\documentclass[useAMS,usenatbib,a4]{mn2e}
\usepackage{graphicx}
\usepackage{amsmath}
\usepackage{amssymb}
\usepackage{amsfonts}
\usepackage{aas_macros}
\usepackage{verbatim}
\usepackage{subfigure}
\usepackage{rotating}
\usepackage{times}

\newcommand{\paczynski}{Paczy{\'n}ski }
\newcommand{\thetae}{\theta_{\mathrm{E}}}
\newcommand{\re}{r_{\mathrm{E}}}
\newcommand{\dl}{D_{\mathrm{l}}}
\newcommand{\ds}{D_{\mathrm{s}}}
\newcommand{\dc}{d_{\mathrm{c}}}
\newcommand{\dw}{d_{\mathrm{w}}}
\newcommand{\tein}{t_{\mathrm{E}}}
\newcommand{\tzero}{t_{\mathrm{0}}}
\newcommand{\thetazero}{\theta_{\mathrm{0}}}
\newcommand{\uzero}{u_{\mathrm{0}}}
\newcommand{\zs}{z_{\mathrm{s}}}
\newcommand{\zone}{z_{\mathrm{1}}}
\newcommand{\ztwo}{z_{\mathrm{2}}}
\newcommand{\mzero}{I_{\mathrm{b}}}
\newcommand{\blendfs}{f_{s}}

\newcommand{\mtwo}{m_{\mathrm{2}}}
\newcommand{\bigmone}{M_{\mathrm{1}}}
\newcommand{\bigmtwo}{M_{\mathrm{2}}}
\newcommand{\zonebar}{\overline{z}_{\mathrm{1}}}
\newcommand{\ztwobar}{\overline{z}_{\mathrm{2}}}
\newcommand{\zbar}{\overline{z}}
\newcommand{\vt}{v_{\mathrm{t}}}
\newcommand{\msun}{M_{\sun}}
\newcommand{\mearth}{M_{\earth}}
\newcommand{\mjup}{M_{\mathrm{Jupiter}}}
\newcommand{\rzero}{R_{\mathrm{0}}}
\newcommand{\ase}[3]{#1_{-#2}^{+#3}}
\newcommand{\dd}{\mathrm{d}}

\newcommand{\mbreak}{M_{\mathrm{break}}}
\newcommand{\vrot}{v_{\mathrm{rot}}}
\newcommand{\vmax}{v_{\mathrm{max}}}
\newcommand{\vl}{v_{\mathrm{l}}}
\newcommand{\vo}{v_{\mathrm{o}}}
\newcommand{\vs}{v_{\mathrm{s}}}
\newcommand{\dzero}{d_{\mathrm{0}}}

\newcommand{\fbs}{\epsilon_{\mathrm{BS}}}
\newcommand{\fom}{\epsilon_{\mathrm{OM}}}
\newcommand{\rv}{R_{\mathrm{v}}}
\newcommand{\rt}{R_{\mathrm{T}}}
\newcommand{\vcirc}{v_{\mathrm{circ}}}
\newcommand{\xs}{x_{\mathrm{s}}}
\newcommand{\ys}{y_{\mathrm{s}}}
\newcommand{\pvec}{\vec{\mathfrak{p}}}

\newcommand{\athresh}{A_{\mathrm{thresh}}}
\newcommand{\tentry}{t_{\mathrm{en}}}
\newcommand{\texit}{t_{\mathrm{ex}}}
\newcommand{\sentry}{s_{\mathrm{en}}}
\newcommand{\sexit}{s_{\mathrm{ex}}}
\newcommand{\single}{^{\mathrm{P}}}
\newcommand{\standard}{^{\mathrm{S}}}
\newcommand{\caustic}{^{\mathrm{C}}}

\newcommand{\tdayunit}{\text{~d}}

\newcommand{\lognormal}{\mathrm{LN}}

\newcommand{\nom}{N_{\mathrm{OM}}}
\newcommand{\nbs}{N_{\mathrm{BS}}}
\newcommand{\nobs}{n_{\mathrm{obs}}}
\newcommand{\nml}{N_{\mathrm{ml}}}

\newcommand{\dcstat}{\Delta\chi_{\mathrm{S}}^2}
\newcommand{\dcsingle}{\Delta\chi_{\mathrm{Pac}}^2}
\newcommand{\dcbase}{\Delta\chi_{\mathrm{b}}^2}
\newcommand{\cbase}{\chi_{\mathrm{b}}^2}
\newcommand{\cstat}{\chi_{\mathrm{S}}^2}
\newcommand{\com}{\chi_{\mathrm{OM}}^2}
\newcommand{\csingle}{\chi_{\mathrm{Pac}}^2}

\title[Orbital Motion in Microlenses] {Detectability of Orbital Motion in
  Stellar Binary and Planetary Microlenses}
\author[M. T. Penny, S. Mao and E. Kerins]
  {Matthew~T.~Penny,$^1$\thanks{Matthew.Penny@manchester.ac.uk} Shude
  Mao,$^{1,2}$ Eamonn Kerins$^1$ \\
$^1$Jodrell Bank Centre for Astrophysics, The Alan Turing Building,
  School of Physics and Astronomy,\\ The University of Manchester,
  Oxford Rd, Manchester, M13 9PL, UK\\
$^2$National Astronomical Observatories, Chinese Academy of Sciences, A20 Datun
Road, Chaoyang District, Beijing 100012, China}
\date{Accepted 2010 October 27.  Received 2010 October 27; in original form 2010 February 10}

\def\LaTeX{L\kern-.36em\raise.3ex\hbox{a}\kern-.15em
    T\kern-.1667em\lower.7ex\hbox{E}\kern-.125emX}

\DeclareGraphicsExtensions{.pdf} 

\begin{document}

\label{firstpage}

\maketitle

\begin{abstract}
A standard binary microlensing event lightcurve allows just two
parameters of the lensing system to be measured: the mass ratio of the
companion to its host, and the projected separation of the components
in units of the Einstein radius. However, other exotic effects can
provide more information about the lensing system. Orbital motion in
the lens is one such effect, which if detected, can be used to
constrain the physical properties of the lens. To determine the
fraction of binary lens lightcurves affected by orbital motion (the detection efficiency) we simulate lightcurves of orbiting binary star and star-planet
(planetary) lenses and simulate the continuous, high-cadence
photometric monitoring that will be conducted by the next generation
of microlensing surveys that are beginning to enter operation. The
effect of orbital motion is measured by fitting simulated lightcurve data with 
standard static binary microlensing models; lightcurves that are poorly
fit by these models are considered to be detections of orbital
motion. We correct for systematic false positive detections by also
fitting the lightcurves of static binary lenses. For a continuous
monitoring survey without intensive follow-up of high magnification events, we find the orbital motion detection efficiency for
planetary events with caustic crossings to be $0.061 \pm 0.010$,
consistent with observational results, and $0.0130 \pm 0.0055$ for
events without caustic crossings (smooth events). Similarly for
stellar binaries, the orbital motion detection efficiency is $0.098
\pm 0.011$ for events with caustic crossings and is $0.048 \pm 0.006$
for smooth events. These result in combined (caustic crossing and
smooth) orbital motion detection efficiencies of $0.029 \pm
0.005$ for planetary lenses and $0.070 \pm 0.006$ for stellar binary
lenses. We also investigate how various microlensing
parameters affect the orbital motion detectability. We find that the
orbital motion detection efficiency increases as the binary mass ratio
and event time-scale increase, and as impact parameter and lens
distance decrease. For planetary caustic crossing events, the
detection efficiency is highest at relatively large values of
semimajor axis~$\sim 4$~AU, due to the large size of the resonant
caustic at this orbital separation. Effects due to the orbital
inclination are small and appear to only significantly affect smooth stellar binary
events. We find that, as suggested by \citet{Gaudi:2009pmc}, many of
the events that show orbital motion can be classified into one of two
classes. The first class, \emph{separational} events, typically show large
effects due to subtle changes in resonant caustics, caused by changes
in the projected binary separation. The second class,
\emph{rotational} events, typically show much smaller effects, which are due to
the magnification patterns of close lenses exhibiting large changes in
angular orientation over the course of an event; these changes
typically cause only subtle changes to the lightcurve.
\end{abstract}


%
%
%
%

%
%
\section{Introduction}

The current gravitational microlensing surveys,
OGLE~\citep{Udalski:2003ews} and MOA~\citep{Hearnshaw:2005moa}
discover $\sim 700$ unique microlensing events per year, of which, of
order ten percent show signatures of lens binarity. A small fraction
of these, those with a high probability of planet detection, are
followed-up by a number of follow-up teams, which intensively monitor
the events for the signatures of planets. In the coming years this
strategy will be augmented and extended by a strategy of continuous, high cadence surveys, performed by a global network of wide field telescopes. Such a network will monitor all the microlensing events it discovers with a
cadence similar to that achieved by the follow-up networks for a
handful of events today. 

The lightcurve of a standard static binary lens, in which the lens
components are fixed and the source follows a straight path, can be
described by a minimum of seven parameters. Only three of these
parameters contain physical information about the lens system. Two are
dimensionless parameters: the mass ratio $q$, and the projected
separation of the lens components $d$, measured in units of the
Einstein radius. The third, the Einstein time-scale of the event
$\tein$, is the time taken for the source to cross one Einstein
radius  
\begin{equation}
\tein=\frac{\re}{\vt},
\end{equation}
where $\vt$ is the relative projected lens-source velocity and $\re$
is the Einstein radius. This is defined as 
\begin{equation}
\re = \sqrt{\frac{4G}{c^2}x(1-x)\ds M},
\label{omrE}
\end{equation}
where $x=\dl/\ds$ is the ratio of the lens distance $\dl$ to the
source distance $\ds$ and $M$ is the total mass of the binary. Of the
other four parameters, three are purely geometrical, and the final
parameter is the unlensed source flux. 

The mass ratio and separation are closely related to the most
interesting properties of the binary, the component masses and the
semimajor axis of the orbit. They can be measured very accurately from
a lightcurve, but only describe the binary's properties in terms of
ratios relative to the typical physical scales of the system. The
Einstein crossing time-scale $\tein$ contains information on these
scales, but this information is wrapped up in a three-fold degeneracy
(the so-called microlensing degeneracy) between the total binary mass,
the lens distance and the source velocity. It is also dependent on the
source distance, but this is usually well constrained by measurements
of the baseline flux. To gain any more knowledge of the lens system
requires that this degeneracy be broken, either by the detection of
higher order effects in the event lightcurve, or by detection of the
lens flux and proper motion as the lens and source
separate.\footnote{Throughout we will use the terms lens motion and
  source motion interchangeably.} These detections yield measurements
of the lens distance and source velocity respectively, allowing the
lens mass to be solved
for~\citep{Gould:1992pmm,Bennett:2006ohs}. Higher order effects, such
as finite source
effects~\citep{Witt:1994fs,Nemiroff:1994mfs,Alcock:1997ffs} and
microlensing
parallax~\citep{Refsdal:1966lmd,Gould:1992mss,Alcock:1995fpe}, allow
the microlensing degeneracy to be broken or reduced through
measurement or constraint of some of the parameters that are combined
in $\tein$. For example detections of finite source effects and
microlensing parallax in the same event yield two independent
measurements of the angular Einstein radius $\thetae=\re/\dl$, which
allow the source velocity and lens distance to be eliminated, and the
lens mass determined~\citep[e.g.][]{An:2001tpe}.

Orbital motion of the binary lens is another such higher order
effect. If the binary lens components are gravitationally bound, they
will orbit each other, and their projected orientation will change as
a microlensing event progresses. As the magnification pattern produced
by a binary lens is not rotationally symmetric, the change in
orientation may be detectable in the lightcurve of the event. If the
orbit is inclined relative to the line of sight, then the projected
separation of the lens components will also evolve, causing changes in
the structure of the magnification pattern, which again may be
detectable. In a small fraction of binary microlensing events we can
expect to see the effects of this orbital motion in their
lightcurves, though this is the first work that attempts to quantify
this fraction. If orbital motion can be detected in a microlens it can
provide constraints on the mass of the lens, and information about the
binary orbit.

To date, six binary microlensing events have shown strong evidence of
orbital motion in the lens system. The first, MACHO-97-BLG-41 was a
stellar mass binary. Modelling of the event was only able to measure
the change in the projected angle and separation of the binary in the
time between two caustic encounters, but was unable to constrain the
orbital parameters~\citep{Albrow:2000rbl}. The second event,
EROS-BLG-2000-5, had very good lightcurve coverage, which allowed the
measurement of the rates of change of the binary's projected
separation and angle; these measurements were then used to obtain a
lower limit of the orbit's semimajor axis and an upper limit on the
combined effect of inclination and
eccentricity~\citep{An:2002eb5}. The third and fourth examples,
OGLE-2003-BLG-267 and OGLE-2003-BLG-291 both seem to show orbital
motion effects~\citep{Jaroszynski:2005bme}. However, only OGLE survey
data was used in their analysis, without follow-up measurements, so
the lightcurve coverage was not ideal. Combined with parallax
measurement, the masses of both binary lenses were constrained, but no
constraints could be placed on the
orbits~\citep{Jaroszynski:2005bme}. In each of these four cases, the
ratio of the component masses is large (near unity), indicative of the
lens systems being binary stars, however, orbital motion has recently
been measured in two events involving planetary mass
secondaries. After the paper was submitted, two further events
  have been shown to display orbital motion effects:
  OGLE-2005-BLG-153~\citep{Hwang:2010vlm} and OGLE-2009-BLG-092/MOA-2009-BLG-137~\citep{Ryu:2010drb}.

OGLE-2006-BLG-109 was an event involving a triple lens, with analogues
of Jupiter and Saturn orbiting a $\sim 0.5 \msun$
star~\citep{Gaudi:2008jsa}. The lightcurve of the event had extremely
good coverage, and showed multiple features, allowing the orbital
motion of the Saturn analogue to be detected. The detection was so
strong that the semimajor axes of both planets could be strongly
constrained~\citep{Gaudi:2008jsa}. A more complete analysis of the
event, incorporating measurements of the lens flux and orbital stability
constraints, carried out by \citet{Bennett:2010jsa}, tightly
constrained four out of six Keplerian orbital parameters of the Saturn
analogue, and weakly constrained a fifth. The planet
OGLE-2005-BLG-071Lb is a $\sim 4$ Jupiter mass planet orbiting a $\sim
0.5 \msun$ star~\citep{Udalski:2005jmp}. Measurements of the orbital
motion in this event have allowed some constraints to be placed on the
planet's orbit~\citep{Dong:2009mmm}. In all six events other higher
order effects have also been detected, most notably microlens parallax
and finite source effects, which are detected in all the events, and
in each case allow the measurement of the lens mass.

Despite these detections, there has been relatively little theoretical
work on orbital motion in microlensing, likely due to the traditional
assumption that the effects of orbital motion on a binary microlens
lightcurve will be small and in most cases
negligible~\citep[e.g.][]{Mao:1991bml}. The problem was first
considered in detail by \citet*{Dominik:1998mrb}, who concluded that
in most microlensing events the effects of lens orbital motion were
likely to be small, though in some cases lightcurves could be
dramatically different. \citet*{Dominik:1998mrb} points out that the
effect is most likely to be seen in long duration binary microlensing
events with small projected binary separations. \citet*{Ioka:1999kre}
also studied the problem, and pointed out that the effect of binary
lens rotation is likely to be important in self-lensing events in the
Magellanic clouds. \citet{Rattenbury:2002hmp} showed that orbital
motion could affect the planetary signatures seen in
high-magnification events.

The six microlensing events that display orbital motion make up a
significant fraction of the few tens of binary microlensing events
that have been
modelled~\citep[e.g.][]{Alcock:2000bml,Jaroszynski:2002obl,Jaroszynski:2004obl,Jaroszynski:2006obl,Skowron:2007obl},
which begins to shed doubt on the previous conclusion that lens
orbital motion is likely to be unimportant in most binary events. The
two planetary events constitute approximately 15 percent of the entire
published microlensing planet population. These observations motivate
us to revisit the question: how likely are we to see lens orbital
motion in a microlensing event? This question is made especially
pertinent in the context of the next generation of high cadence
microlensing surveys which will make the exquisite lightcurve coverage
of EROS-BLG-2000-5 and OGLE-2006-BLG-109 the norm rather than the
exception. To gain a better understanding of how frequently orbital
motion affects microlensing lightcurves we simulate a large number of
microlensing events caused by orbiting binary lenses. We also
investigate the factors that affect this frequency. 

The structure of the work is as follows. In Section~\ref{MLWithOB}
we will review the basic theory of binary microlensing, and the
effects of orbital motion on such lensing
systems. Section~\ref{Method} describes our simulations of
microlensing events and Section~\ref{MeasOrbMotion} describes how we
measure the effects of orbital motion. In Section~\ref{Results} we
present the results of the simulations. We discuss the results in
Section~\ref{Discussion} and conclude in Section~\ref{Summary}. 

%
%
\section{Microlensing with orbiting binaries}
\label{MLWithOB}

\subsection{Binary microlensing}
\label{binaryMicrolensing}

The lens equation of a binary point mass gravitational lens describes
the mapping of light rays from the source plane to the image plane,
and can be written in complex form~\citep{Witt:1990blc}
\begin{equation}
\zs = z - \frac{1}{1+q}\left(\frac{1}{\zbar-\zonebar} +
  \frac{q}{\zbar-\ztwobar}\right)
\label{LEGeneral}
\end{equation}
where $\zs=\xs+i\ys$ is the complex coordinate
in the source plane, $z=x+iy$ is the complex
coordinate\footnote{The symbol $x$ used here should not be confused
  with that representing the ratio of lens to source distances. Its
  meaning should be clear from the context in which it is used, and it
  will not be used again in this context without a subscript.} in the
image plane, $q\le 1$ is the mass ratio of the secondary mass to the
primary, $\zone=x_{\mathrm{1}}+iy_{\mathrm{1}}$ and $\ztwo =
x_{\mathrm{2}} + iy_{\mathrm{2}}$ are the complex coordinates of the
primary and secondary lens respectively, and bars represent complex
conjugation. All lengths have been normalized to the Einstein radius
of the total lensing mass. The positions of the images $z$ for a given
source position are found by solving this equation, which can be
rearranged into a fifth-order complex polynomial in $z$. The total
magnification of the images is given by the ratio of image areas to
the source area. This information is contained in the Jacobian of the
lens mapping $J$, and the magnification is given by
\begin{equation}
A = \frac{1}{\det J},
\label{Magnification}
\end{equation}
where $\det J$ is the determinant of the Jacobian and is
given by
\begin{equation}
\det J = 1 - \left|\frac{1}{1+q}\left( \frac{1}{(\zbar-\zonebar)^2} +
    \frac{q}{(\zbar-\ztwobar)^2}\right)\right|^2.\label{detJacobian}
\end{equation}

It is possible for $\det J$ to be zero; if this occurs, the
magnification of a point source will be infinite at the point in the
source plane where $\det J=0$. This occurs when the
quantity within the modulus sign in equation~(\ref{detJacobian}) lies on
the unit circle in the Argand diagram. Any point in the image plane
that obeys $\det J=0$ is a critical point. The set of
critical points form a set of closed curves called critical
curves. The critical curves can be found by solving the critical curve
equation
\begin{equation}
\frac{1}{\left(z-\zone\right)^2} + \frac{q}{\left(z-\ztwo\right)^2} =
(1+q)e^{i\phi}
\label{CritCurves}
\end{equation}
where $\phi$ is a parameter, such that when swept over
$0\le\phi<2\pi$, the solutions $z$ draw out the critical curves. These
can then be mapped through the lens equation~(\ref{LEGeneral}) onto
the source plane to form caustics. A point source that lies on such a
caustic will be infinitely magnified, but this unphysical
magnification remains finite for physical, finite sources. The
caustics are characteristically made up of smooth curves, fold
caustics, that meet at cusps.

\begin{figure}
\includegraphics[width=84mm]{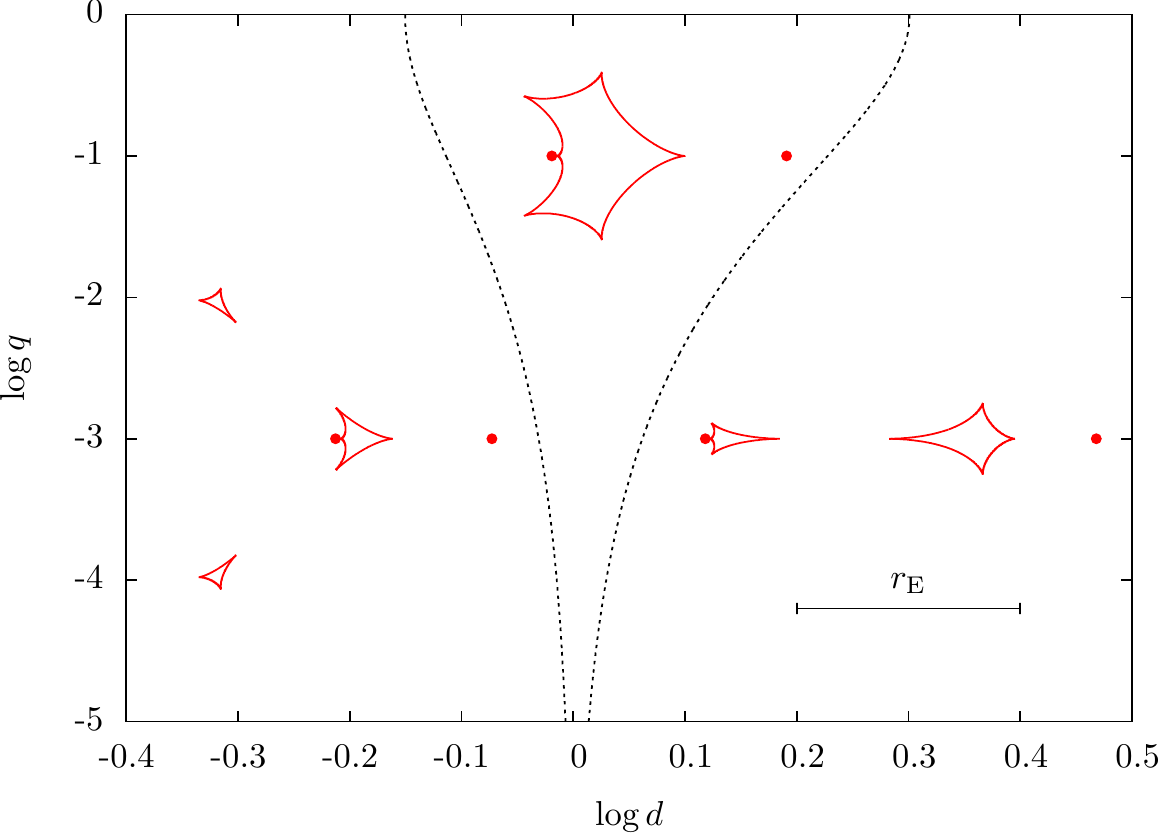}
\caption{Plot showing the $d$-$q$ plane separated into three regions
  where the caustics take on close, resonant and wide topologies,
  with increasing $d$; the dashed lines, $\dc(q)$ and $\dw(q)$, separate the regions of different topology. The solid lines are examples
  of each of the caustic topologies, drawn to the same scale, for a
  binary lens with $q=0.1$, and $d=0.7$ (close), $d=1.05$ (resonant)
  and $d=1.75$ (wide). The filled circles show the positions of
  the lenses for each topology, with the primary (more massive) lens
  being positioned leftmost in all cases. After \citet{Cassan:2008apc}.}
\label{causticTopology}
\end{figure}

The number and shape of caustics is determined by just two lens
parameters, the mass ratio $q$ and the projected lens separation $d$
in units of Einstein radii. There are three possible topologies that
the caustics can assume: close, resonant and wide; examples of each
are shown in Figure~\ref{causticTopology}. The lines that delimit the
different topologies in the $(d,q)$ plane, also shown in the figure,
are given by~\citet{Erdl:1993ctc}: 
\begin{equation}
d_{\mathrm{c}}^8 = \frac{(1+q)^2}{27q}(1-d_{\mathrm{c}}^4)^3,
\label{dc}
\end{equation}
which separates regions of close and resonant topology, and
\begin{equation}
d_{\mathrm{w}}^2 = \frac{(1+q^{\frac{1}{3}})^3}{1+q},
\label{dw}
\end{equation}
which separates regions of resonant and wide topology.

\subsection{Orbital motion in a binary microlens}
\label{orbitalMotion}

The lightcurve of a microlensing event can be considered as a
one-dimensional probe, by the source, of the two-dimensional
magnification pattern produced by the lens. The magnification pattern
of a single lens is rotationally symmetric about the position of the
lens, but the magnification pattern of a binary lens is more
complicated, containing strong caustic structures that exhibit a
reflectional symmetry about the binary axis, the axis connecting the
lens components. However, far away from the caustics, the
magnification pattern can resemble that of a single lens.

As the lens components orbit each other, their position angle and
their projected separation can change. These changes cause changes in
the orientation and structure of the magnification pattern
respectively. It is clear, however, that only if the source traverses
regions of the magnification pattern that differ significantly from
that of a single lens, will it be possible to detect these effects of
orbital motion. For the effects to be measurable the lightcurve of
the event must be affected in a significant way, that is not
reproducible by a static binary lens model. It is also possible to
detect the effect of orbital motion by showing that a static model is
less physically plausible than an orbiting model, but this will
usually require further information about the event, such as an independent
constraint on the lens mass.

The effects of orbital motion on a lightcurve can also be mimicked by
other higher order effects, especially parallax and xallarap. Parallax
effects are caused by the motion of the earth about the sun, and cause
the source to take an apparently curved path through the magnification
pattern~\citep*[e.g.][]{Smith:2003ap}. In the case of xallarap, the
source travels along a curved path through the magnification pattern
as a result of binary orbital motion in the source
system~\citep{Griest:1992bsm, Paczynski:1997, Dominik:1998mrb,
  Rahvar:2009mxe}. These curved paths can look very similar to those
taken by the source in the rotating binary lens centre of mass frame,
and hence it can sometimes be difficult to identify the true cause of
the effect.  

%
%
\section{Simulating a high cadence microlensing survey}
\label{Method} 

The major aims of this study are two-fold: firstly to determine the
fraction of microlensing events that will be affected by orbital
motion, as seen by the next generation microlensing surveys; and
secondly, to investigate the factors that affect the detectability of
orbital motion, to aid the targeting of such events without resorting
to exhaustive modelling efforts. To achieve the first goal, the
various factors that go into the observation of a microlensing event
should be simulated, accurately modelling the observing setup, the
distributions of planetary and binary star lens systems,
and the distribution of the sources and lenses throughout the
Galaxy. To achieve the second goal we must simplify the parameter
space we investigate as far as possible, without removing essential
elements from the model, so as to allow a clear interpretation of the
results.

To balance these somewhat contradictory requirements we choose to
accurately simulate ideal photometry and use a semi-realistic model
of the Galaxy, while investigating a logarithmic distribution of
companion masses and separations. This allows us to use our
simulations to gain a good order of magnitude estimate of the results
expected from future surveys, whilst simultaneously investigating the
factors that have the largest impact on the detection of orbital
motion over a relatively uniform parameter space. 

\subsection{The Galactic model}
\label{GalacticModel}

To simulate the kinematic and distance distributions of the source and
lens populations we assume a simplistic bulge and disk model of the
Galaxy. We assume the source to be located in the bulge, at a fixed
distance $\ds=\rzero=8$~kpc, in the direction of Baade's Window, where
$\rzero$ is the distance to the Galactic centre. The lens
distances are distributed according to the stellar density
distribution of Model II of \citet*{Binney:2008gd}, which consists of
a thin and a thick exponential disk and an oblate spheroidal bulge
with a truncated power-law density distribution. The kinematics of our
Galactic model are based on that of \citet*{Han:1995mpm}, who describe
the kinematics of a stellar disk and a barred bulge. The distribution
of relative source velocities $\dd n/\dd\vt$ is dependent on the
transverse velocities of the lens, source and observer, and their
corresponding velocity distributions. The observer is assumed to
follow the Galactic rotation at the position of the Sun, and therefore
has a velocity
$(v_{\mathrm{O},\ell},v_{\mathrm{O},b})=(225.2,7.2)$~km~s$^{-1}$ in the
directions of Galactic coordinates $(\ell,b)$, once the Solar apex
motion is included. The source and lens are assumed to follow the
Galactic rotation, with an additional random component. Their
velocities have the form, in the directions $\ell$ and $b$, 
\begin{equation}
v_{\ell} = \vrot + v_{\mathrm{rand},\ell}\text{, } v_b = v_{\mathrm{rand},b},
\label{lsvelocities}
\end{equation}
where $\vrot$ is the rotational component of the velocity, and
$v_{\mathrm{rand},\ell}$ and $v_{\mathrm{rand},b}$ are random velocities
in the directions $\ell$ and $b$ respectively.  The rotation curve of
the bulge is assumed to be flat beyond a distance of $1$~kpc from the
Galactic centre, and that of a solid
body within $1$~kpc. Therefore, the rotational velocity component
$\vrot$ for bulge stars is
\begin{equation}
\vrot = \left\{ \begin{array}{c}\vmax \left(
      \frac{R}{\text{kpc}} \right) \text{ if }
    R<1 \text{ kpc}  \\
\vmax \text{ if } R\ge 1
\text{ kpc} \end{array}\right.\label{rotationalVelocity}
\end{equation}
where $\vmax=100\text{~km~s}^{-1}$ is the maximum rotational
velocity of the bulge, and $R=\sqrt{X^2+Y^2}$ and $(X,Y,Z)$ is a
galactocentric coordinate system with the $X$-axis increasing towards
the observer and the $Z$-axis pointing out of the Galactic plane; for
the disk $\vrot=200\text{~km~s}^{-1}$. The random velocity components
are assumed to follow Gaussian distributions, with dispersions taken
from \citet*{Han:1995mod}. These dispersions are $(\sigma_{\ell},\sigma_b)
= (30,20)\text{~km~s}^{-1}$ for the disk and
$(\sigma_X,\sigma_Y,\sigma_Z) = (110,82.5,66.3)\text{~km~s}^{-1}$ for
the bulge. From these quantities, the relative transverse velocity of the
source $\vt$ (the quantity we are interested in) can be calculated
from the relative velocities in the $\ell$ and $b$ directions $v_{\ell}$ and
$v_b$ as
\begin{equation}
\vt = \sqrt{v_{\ell}^2 + v_b^2},
\label{SourceVelocity}
\end{equation}
where~\citep[e.g.][]{Han:1995mpm}
\begin{equation}
v_{\ell,b} = (\vl-\vo)_{\ell,b} + x(\vo-\vs)_{\ell,b},
\label{vl}
\end{equation}
and $\vo,\vl$ and $\vs$ are the observer, lens and source velocities
respectively, in the directions $\ell$ and $b$.

\begin{figure}
\includegraphics[width=84mm]{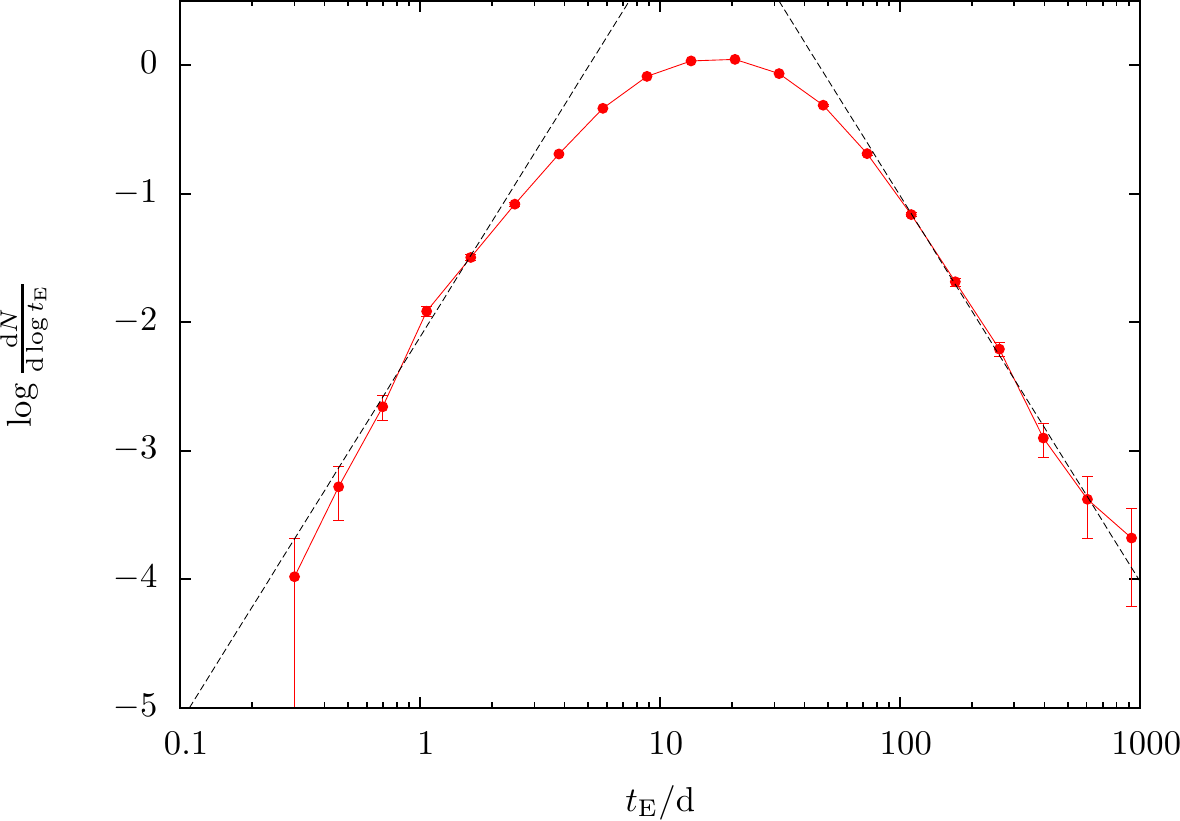}
\caption{The Einstein time-scale distribution for $\sim 50000$ simulated
  events. The solid line and data points show the simulated data, and
  the dashed lines show lines of slope $3$ and $-3$, the expected asymptotic behaviour of the distribution.}
\label{tEDistribution}
\end{figure}

The final distribution of lens distances and velocities takes into
account the dependence of the event rate $\Gamma \propto \sqrt{x(1-x)}\vt$
on the distribution of each parameter. While the kinematic and density
distributions are produced from different Galactic models, they
qualitatively reproduce the observed Einstein crossing time-scale
distribution, shown in Figure~\ref{tEDistribution}, including its
asymptotic behaviour~\citep{Mao:1996mm}.

\subsection{The microlensing events}
\label{lensSystems}

When observing a microlensing event, it is often the case that the
light of the source being magnified is blended with that of nearby
stars in the field. The amount of blending can be quantified by a
blending fraction $\blendfs$, which we define to be the fraction of the
total flux of the observed blend that the source contributes when
unmagnified, such that the time dependent magnitude of the blend is
\begin{equation}
I(t) = \mzero - 2.5\log(\blendfs A(t) + (1-\blendfs)),
\label{blendedMagnitude}
\end{equation}
where $\mzero$ is the baseline magnitude of the observed blend when
the source is unmagnified, and $A(t)$ is the magnification caused by
the lens.

The distribution of baseline magnitudes and blending fractions is
drawn from simulations of blending effects by \citet{Smith:2007bl},
who perform photometry on mock images of typical Galactic bulge fields
with high stellar density. Specifically we calculate the blending
fraction and baseline magnitude for each event from the input and
output magnitudes of source stars drawn from their simulation with
$1.05$~arcsec seeing and input stellar density of
$133.1$~stars~arcmin$^{-2}$, before any detection efficiency cuts are
made to the catalogue. As the phenomenon of negative
blending, the source apparently contributing a fraction $\blendfs>1$
to the total flux of the blend~\citep{Park:2004jpd,Smith:2007bl}, is
poorly understood, we only include sources with moderate negative
blending, requiring that $\blendfs<1.2$.

The mock images are produced by \citet{Smith:2007bl} using the method
of \citet{Sumi:2006odm}, drawing stars from the Hubble Space Telescope
$I$-band luminosity function of \citet{Holtzman:1998blf}, adjusted to
account for denser fields and brighter stars using OGLE
data. Extinction was accounted for using the extinction maps of
\citet{Sumi:2004bem}, and the baseline magnitudes were measured using
the standard OGLE pipeline based on {\sc
  dophot}~\citep{Schechter:1993dop}. Further details can be found in
section~3 of the \citet{Smith:2007bl} paper, and references therein.

The lens systems are composed of a primary of mass $\bigmone$,
and secondary of mass $\bigmtwo$. The primary's mass is drawn from a
broken power-law distribution
\begin{equation}
\frac{\dd n}{\dd\bigmone}\propto \bigmone^{(\alpha+0.5)};\text{ }
\alpha=\left\{\begin{array}{cr} -1.3 & \bigmone\leq\mbreak\\-2.0 &
\bigmone>\mbreak\end{array}\right.,
\label{massFunction}
\end{equation}
with lower and upper limits of $0.05\msun$ and $1.2\msun$
respectively, and where $\mbreak=0.5\msun$. The addition of $0.5$ to
the power-law index is to account for the dependence of the
microlensing event rate on the mass of the lens. We do not include a
population of stellar remnant lenses, such as white dwarfs, neutron
stars and black holes. The mass ratio $q$ of the secondary to the
primary is drawn from a logarithmic distribution, with limits
$10^{-2}\le q < 1$ for stellar binary lenses and $10^{-5}\le q <
10^{-2}$ for planetary lenses. Note that for lower mass primaries, the
distribution of stellar binary mass ratios does include secondaries
with masses as low as $\sim 5\mjup$, well into the planetary mass
regime, and the lower limit of the planetary mass ratio distribution
implies a secondary of $\sim 1\mearth$ for a $0.3\msun$ primary.

The components of the lens orbit their combined centre of mass in
Keplerian orbits, of semimajor axis $a$, distributed logarithmically
over the range $0.1\text{AU}\leq a<20\text{AU}$. These orbits are
inclined to the line of sight, with inclination angles distributed
uniformly over a sphere. For binary stars we performed two sets of simulations, one with zero eccentricity $e$, and another with bound, eccentric
orbits with eccentricities distributed uniformly over $0\le e < 1$.

The source trajectories were parametrized by the angle of the source
trajectory relative to the binary axis $\thetazero$, at the time of
closest approach $\tzero$, and the impact parameter $\uzero$, the
projected source-lens separation in units of Einstein radii at
$\tzero$. We set $\tzero=0$, for simplicity, and $\thetazero$ and
$\uzero$ were distributed uniformly over the ranges
$0\le\thetazero<2\pi$ and $-1.5\le\uzero<1.5$ respectively.

\subsection{Simulation of photometry}
\label{photometry}

In the hunt for planets, the proposed next generation of microlensing
survey will consist of a (potentially homogeneous) network of
telescopes located throughout the southern hemisphere such that the
target fields in the Galactic bulge can be monitored continuously
during the times when the bulge is observable. The telescopes will
have diameters between $1.3$--$2.0$~m and fields of view
$1.4$--$4.0$~square degrees. They will operate at a cadence of
approximately 10~minutes, and are expected to discover several
thousand microlensing events per year. An example is KMTNet, a
network of three identical $1.6$~m telescopes due to enter operation
in 2014~\citep{Kim:2010kmt}. Such surveys can operate effectively
without the need for intensive follow-up observations due to their
high cadence and continuous coverage. However, it is likely that the
survey/follow-up observing paradigm will persist, with low cadence
surveys monitoring far larger areas of sky.

Unfortunately the effects of the weather, amongst other things, makes completely continuous, high-cadence observations unachievable in reality. Rather than including complicated models of these effects, as well as other
effects such as the lunar cycle and their effects on the photometry, we instead choose a simpler prescription. Each event is monitored with continuous photometry at a reduced cadence of 30~minutes. These observations are performed by telescopes with $1.3$~m effective diameter observing in the $I$-band. For each exposure of $120$~s, the seeing is chosen from a lognormal distribution with mean $1.2$~arcsec and standard deviation $0.25$~arcsec, and a background flux distributed as
\begin{equation}
F = 8500 \lognormal (1.5,0.4)~\text{photons~arcsec}^{-2},
\label{backgroundFlux}
\end{equation}
which is integrated over a seeing disk, and where
$\lognormal(\mu,\sigma)$ is a lognormal distribution with mean $\mu$
and standard deviation $\sigma$. New values of seeing and background
flux are chosen for each observation. A lower limit on the photometric
accuracy is imposed by adding a Gaussian noise component, with
dispersion 0.3~percent, to the photon counts, which are calculated by
assuming 10~photons~m$^{-2}$~s$^{-1}$ reach the observer from a $I=22$
source.

\begin{figure}
\includegraphics[width=84mm]{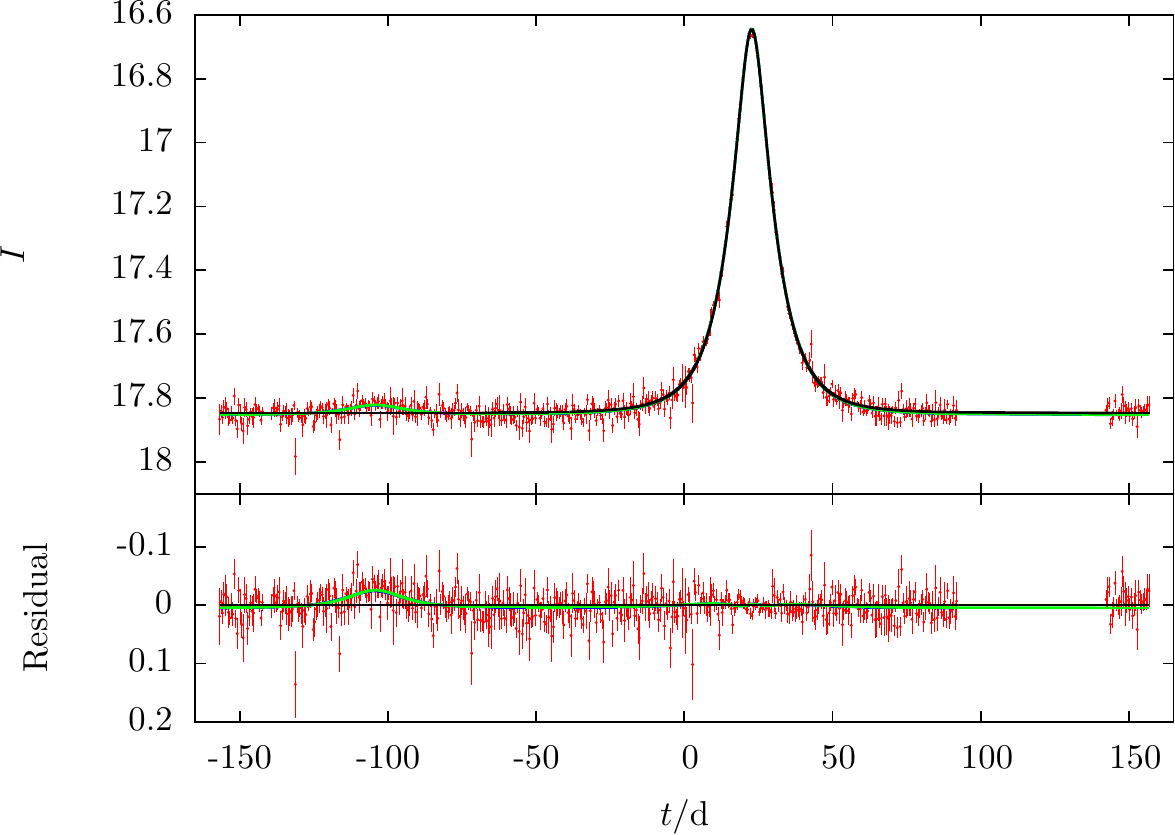}
\caption{An example lightcurve from the simulations that required
  coverage to be extended to cover a feature far from the lightcurve
  peak. The upper panel shows the lightcurve ($I$-band magnitude) and
  the lower panel shows the \paczynski residual ($I$-band residuals
  from the single lens fit). Red points show the simulated data points
  with error bars, and black, green and blue lines are the
  best-fitting \paczynski model, the best-fitting static binary model
  and the true orbital motion model (largely hidden below the green
  static model curve) respectively. Only one data point in 24 is shown
  for clarity. The lightcurve shown is for that of an event by a
  stellar binary lens with $q = 0.22$, $d \approx 8.6$ and $\tein = 14.9
  \tdayunit$. Usually, only data points that cover the inner $5\tein$
  are used, apart from some data points used to constrain the baseline
  magnitude (cf. the lightcurve for times $t > 0\tdayunit$), however
  additional data points are used to fully cover the additional
  lightcurve feature down to baseline (cf. the lightcurve for $t <
  0\tdayunit$). Further details for the event can be found in Tables~4
  and 5 in the online supplementary material.}
\label{lightcurveCoverage}
\end{figure}

To ensure that all the features of a lightcurve are covered, and that
there is a good balance between the baseline, peak and features of the
lightcurve when fitting (see the next section), the lightcurve is
monitored continuously over the times $-5\tein \le t-\tzero < 5\tein$,
and over $10.5\tein\leq|t-\tzero|<9.5\tein$ to sample the baseline. To
ensure that all features are covered, if the magnification of the
source rises above $A\geq \athresh=1.0062$, the coverage is extended
so as to be continuous within one Einstein time-scale of the feature and
continuous between the feature and
$t=\tzero$. Figure~\ref{lightcurveCoverage} shows an example of a
lightcurve where coverage had to be extended.

\section{Measuring orbital motion}
\label{MeasOrbMotion}

Ultimately we are interested in finding the fraction of binary
microlensing events which show signs of orbital motion. To do this we
must classify the events we simulate into those binary events that do
show orbital motion, those that do not, and events that do not show
binary signatures. To do this we fit each event first with a single
lens model, and then those events which are poorly fit with this
model, we fit with a static binary lens model. To evaluate the
effectiveness of each stage of the fitting process, in addition to the
fitting of the lightcurves simulated with orbiting binary lenses, we
must also fit lightcurves simulated with point-mass lenses and static
binary lenses. 

\subsection{Lightcurve modelling}

The single lens model has five parameters: the time of closest
approach $\tzero\single$, the event time-scale
$\tein\single$, the impact parameter $\uzero\single$, the
baseline magnitude $\mzero\single$ and the blending fraction
$\blendfs\single$. We performed a $\chi^2$ minimization using the {\sc
  minuit} routine from {\sc cernlib}~\citep{minuit}, with all
parameters free; all parameters were unconstrained, except for
$\blendfs\single$, which was constrained to be within
$0.0<\blendfs\single<1.2$. For each event, seven single lens fits were performed,
with different initial blending fractions, 
$\blendfs\single=0.05$, $0.2$, $0.4$, $0.6$, $0.8$, $1.0$ and
$1.2$. For each fit, the initial guesses for each parameter were:
$\tzero\single=0$, the time-scale was the true time-scale, the
baseline magnitude was taken to be the magnitude of the first data
point on the lightcurve, and the impact parameter was chosen such that
at $t=\tzero\single$ the magnitude of the event would be that of
the brightest data point. This prescription works well for
events which are well modelled by a single lens model, but not so for
events with strong binary features, or events which are heavily
blended and barely rise above baseline. It is therefore useful to
eliminate events falling into the later category before performing the
fitting, such that the only events that the single-lens model fails to
fit are ones that show genuine signs of lens binarity. This cut will
be described in the next subsection.

To fit the binary lens lightcurves, we found it necessary to split the
events into caustic-crossing events and non-caustic-crossing events,
and to fit each category using a different parametrization. The
non-caustic crossing events we fitted with a standard parametrization,
with a reference frame centred on the primary lens. The parameters are:
the time of closest approach to the lens primary
$\tzero\standard$, the event time-scale $\tein\standard$, the
impact parameter between the lens primary and the source
$\uzero\standard$, the angle of the source trajectory to the
binary axis $\thetazero\standard$, the logarithm of the projected
binary separation $\log d\standard$, the logarithm of the normalized
secondary mass $\log \mtwo\standard$, the baseline magnitude
$\mzero\standard$ and the blending fraction $\blendfs\standard$. For
brevity we introduce the vector notation
\begin{equation}
\pvec\standard = \left( \tzero\standard, \tein\standard,
  \uzero\standard, \thetazero\standard, \log d\standard, \log
  \mtwo\standard, \mzero\standard, \blendfs\standard \right),
\label{standardParameters}
\end{equation}
to represent the parameter set of the standard binary
parametrization. 

For the number of lightcurves necessary to obtain a good statistical
sample, a full search of the full binary lens parameter space is not
computationally feasible, so we perform just one minimization per
lightcurve. We must therefore pay special attention to the choice of 
initial guesses we use, firstly so as to maximize the chance of
finding a good minimum, and secondly so as to treat the fitting of
the static binary events comparably to the orbiting binary events. The
static binary simulations are drawn from the same distributions as the
orbiting binary simulations, the only difference being that the lens
is frozen in the state it would be in at $t=\tzero$.

As we have simulated the microlensing events, we already have a
perfect knowledge of the systems, and we can use this knowledge to
obtain a good set of initial guesses. We note that at a given time,
the state of an orbiting binary lens can be described by a static
binary model. We can therefore describe our lens at time $t$
using the time dependent parameter set 
\begin{equation}
\pvec(t) = \left( \tzero, \tein, \uzero, \thetazero(t), d(t), q,
  \mzero, \blendfs \right),
\label{timeDependence}
\end{equation}
where we have used the same definitions and centre of mass reference
frame as in the previous section. Note that only two of the parameters
are time dependent, and so we can use the true values of the constant
parameters as initial guesses, having applied the appropriate
coordinate transformations.\footnote{In the reference frame of
  $\pvec\standard$, $\tzero$ and $\uzero$ would also be time dependent
  as the origin (the primary mass) is not fixed.} However, we are
still left with the problem of choosing the guesses of
$\thetazero\standard$ and $d\standard$. We could choose
$\thetazero(\tzero)$ and $d(\tzero)$, but this would bias the fitting
success probability unfairly towards static binary events: the initial
guess would be the actual model used to simulate the data. Instead, we
choose to use $d(t_{\mathrm{f}})$ and $\thetazero(t_{\mathrm{f}})$,
where $t_{\mathrm{f}}$ is the time of a feature in the lightcurve. We
define a feature simply as any maximum in the lightcurve, or a maximum
or minimum in the \paczynski residual (the residual of the true
lightcurve with respect to the best fitting single lens model) with
$|I-I_{\mathrm{Pac}}|>0.1$, where $I$ is the $I$-band magnitude of the
true model, and $I_{\mathrm{Pac}}$ the $I$-band magnitude of the best
fitting \paczynski model. As there are in general more than one feature, we
choose the feature that gives the best
$\chi^2(\pvec(t_{\mathrm{f}}))$. If the initial guesses for fits to
static binary lightcurves are chosen in the same way, as if the binary
were orbiting, then the initial guesses for static lenses should be
worse than for orbiting lenses, as at the time of the chosen feature,
the true orbiting lens magnification will exactly match the
magnification of the initial guess static model. In reality, for
$t_{\mathrm{f}}\approx\tzero$ there will likely be a bias in favour of
static lenses and $t_{\mathrm{f}}\not\approx\tzero$ there will be a
bias in favour orbiting lenses, but we do not believe this will affect
results significantly. To fit the events, we again use the {\sc minuit}
minimizer, allowing all parameters to vary. All parameters are
unconstrained, except for $\blendfs\standard$, which is constrained to
the range $0<\blendfs\standard<1.2$. 

While this method was suitable for events which showed smooth binary
features, it is not always suitable for those events which exhibit caustic
crossings. For these events, in addition to fitting with the standard
parametrization, we also used the alternative parametrization of
\citet{Cassan:2008apc}. This replaces the parameters specifying the
source trajectory $(\tzero\standard, \tein\standard, \uzero\standard,
\thetazero\standard)$, with parameters that better reflect the sharp
caustic crossing features of the lightcurve $(\tentry\caustic,
\texit\caustic, \sentry\caustic, \sexit\caustic)$ the times of a
caustic entry and exit and the positions of the entry and exit on the
caustic respectively, where $\sentry\caustic$ and $\sexit\caustic$,
are defined to be the chord length along the caustic, normalized such that
$0\le\sentry\caustic,\sexit\caustic<2$. Full details of the
parametrization can be found in the \citet{Cassan:2008apc} paper. The
parameter set we use for caustic-crossing events is therefore
\begin{equation}
\pvec\caustic = \left( \tentry\caustic, \texit\caustic,
  \sentry\caustic, \sexit\caustic, \log d\caustic, \log q\caustic,
  \mzero\caustic, \blendfs\caustic \right),
\label{causticParameters}
\end{equation}
where the parameter $\log \mtwo\standard$ has been replaced by
$\log q\caustic$ as a matter of preference; the two parameters
are related by $\mtwo\standard = q\caustic/(1+q\caustic)$.

\begin{figure}
\includegraphics[width=84mm]{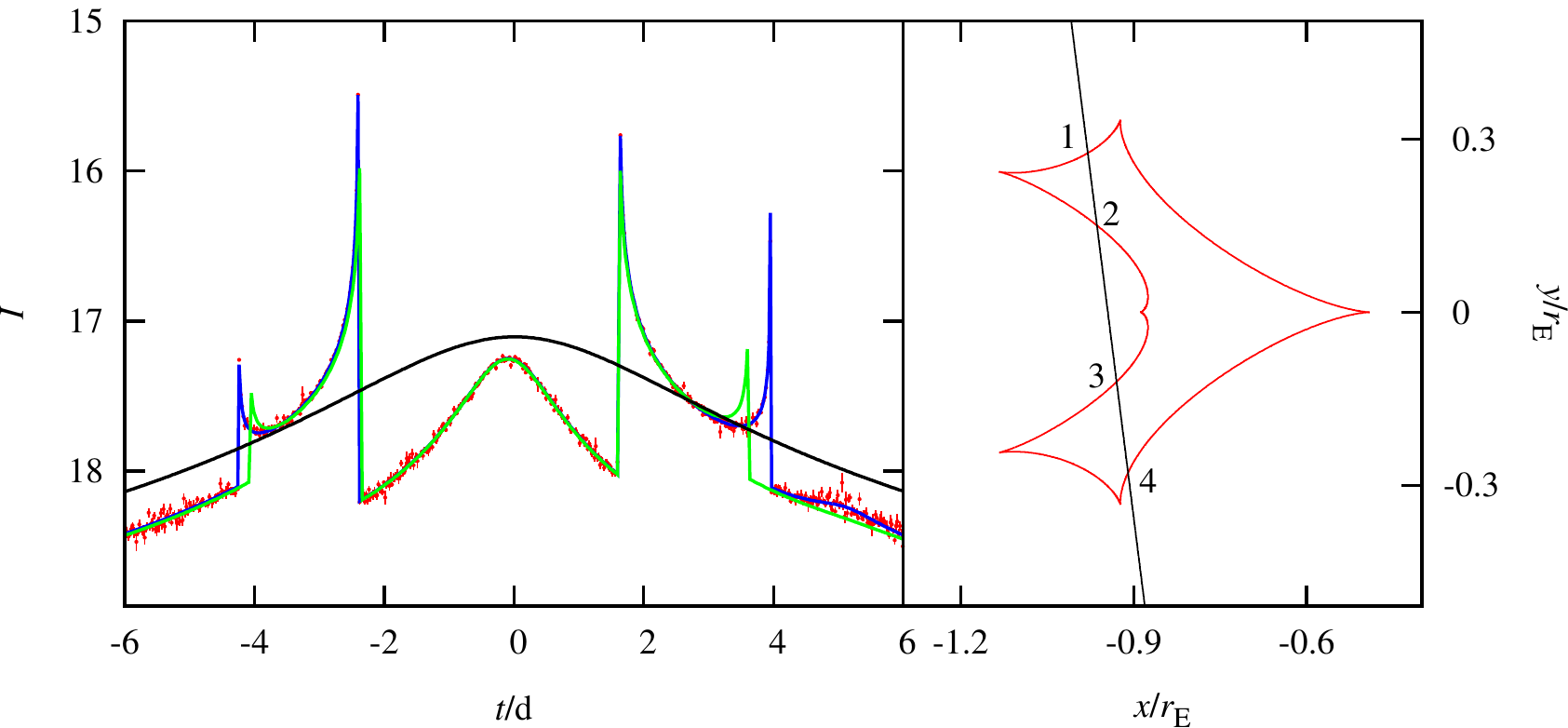}
\caption{Example lightcurve and caustic map of an event where a
  non-adjacent caustic entry-exit pair was chosen for fitting with the
  \citet{Cassan:2008apc} parametrization. The lightcurve is shown in
  the left panel, where red points show the simulated data, the blue
  line is the true model and the green line is the static binary
  model. The right panel shows a map of the caustic of the static
  binary model, plotted in red, and the source trajectory, plotted in
  black. The numbers indicate the order of the caustic crossings. The
  static model has been adjusted by hand to aid clarity. Further
  details for the event can be found in Tables~4 and 5 in the online
  supplementary material.}
\label{multipleCrossings}
\end{figure}

The accurate calculation of the $\sentry\caustic$ and $\sexit\caustic$
parameters is quite computationally expensive, compared to the
calculation of a lightcurve, and needs to be repeated each time $d$ or
$q$ changes. Also, despite the improved parametrization, the $\chi^2$
surface is still very complicated, especially in the
$\sentry\caustic$-$\sexit\caustic$ plane, containing many local
minima. For these reasons we pursue a three stage minimization
process. We begin by conducting a grid search over the entire
$\sentry\caustic$-$\sexit\caustic$ plane, with $128 \times 128$ points
spaced evenly in $\sentry\caustic, \sexit\caustic$, and with all other
parameters, including the caustic crossing times, fixed at their true
values, except for $\log d\caustic$. $\log d\caustic$ is fixed at a
random value is chosen in from the range $\Delta\log d\caustic =
1.5(\log d(\texit) - \log d(\tentry))$ or $\Delta\log d\caustic =
0.015$, whichever is greater, centred on the midpoint of $\log d$
between the caustic crossings, and where $d(\tentry)$ and $d(\texit)$
are the projected separations at the caustic entry and exit times
respectively. The range of $\Delta\log d\caustic$ is truncated, if
necessary, to ensure that it only covers the caustic topologies at the time of the crossings. For the
static lenses, $\log d\caustic$ is chosen from a uniform distribution with the
same range as if the lens were orbiting. The grid search is then refined by performing a second
$128 \times 128$ grid search over a box of side length $1/32$ about
the grid point with the lowest $\chi^2$. Six $2\times128\times128$
grid searches are performed with different random values of $\log
d\caustic$. In cases where there are multiple caustic crossings,
different pairs of caustic crossings are used to define $(\tentry\caustic,
\texit\caustic, \sentry\caustic, \sexit\caustic)$ for each grid
search. Figure~\ref{multipleCrossings} shows an example lightcurve
where the first caustic exit defines $(\tentry\caustic,
\sentry\caustic)$ and the second caustic entry defines $(\texit\caustic,
\sexit\caustic)$.

The second stage of the fitting simply polishes the result of the
grid search by performing a minimization over $\sentry\caustic$ and
$\sexit\caustic$, with all other parameters fixed, using {\sc
  minuit}. In the final stage of the fitting, all parameters, except
for $\tentry\caustic$ and $\texit\caustic$ are allowed to vary in a
further {\sc minuit} minimization. Again, all parameters were
unconstrained, except for $\blendfs\caustic$ which was constrained to
the range $0 < \blendfs\caustic < 1.2$. We found that, at all stages of
the minimization for caustic crossing events, the
minimization performed better when the first and last data points
inside the caustic crossing were not considered in the
fit. This is because, with the high cadence observations that we
simulate, the point source is typically very close to the inside of
the fold caustic, and hence is magnified by many orders of
magnitude. This leads to unrealistic photometry in two ways: firstly,
in a real detector, saturation would become a problem, and secondly, a
real, finite, source would not be magnified in such an extreme
way.

\subsection{Classification of events}

With the modelling procedures in place, we now describe the
classification of the events. The classification is performed by a
series of cuts based on the $\chi^2$ results of the fitting described
in the last subsection. The first cut, the variability cut, removes
events which do not show significant variability from the
analysis. This is done, without fitting, by comparing the $\chi^2$
values of the simulated data relative to the true model, $\com$, and
relative to a constant lightcurve with no variability at the true baseline
magnitude, $\cbase$. We exclude events that do not satisfy
\begin{equation}
\frac{\dcbase}{\nobs} \equiv \frac{\cbase - \com}{\nobs} > 0.3,
\label{constCut}
\end{equation}
where $\nobs$ is the number of observations.

The second cut is used to classify events into single lens-like
events, and binary lens-like events, or events that do not and do
exhibit binary lens features in their lightcurves. Using the results
of the single lens modelling, $\csingle$, the $\chi^2$ of the
simulated data with respect to the single lens model, we define events
that satisfy
\begin{equation}
\dcsingle \equiv \csingle - \com > 200
\label{singleCut}
\end{equation}
to be binary events, and those that do not to be single events. Binary
events can then be split into caustic crossing binary events and
smoothly varying events, or caustic crossing and smooth events
respectively. We define a caustic crossing event as one where at least
one data point is measured when the source is inside a
caustic.\footnote{The removal of data points in the fitting process
  does not affect the classification.}

The final cut is based on the result of lightcurve fitting with binary models. Events that satisfy
\begin{equation}
\dcstat \equiv \cstat - \com > 200
\label{binaryCut}
\end{equation}
are classified as events that exhibit orbital motion (orbital motion
events) and those that do not are classified as static events, where
$\cstat$ is taken to be the $\chi^2$ of the best fitting static binary
model. For smooth events this is the $\chi^2$ of the best fitting standard
binary model, and for caustic crossing events it is the $\chi^2$ of
the better fitting of the \citet{Cassan:2008apc} caustic crossing
model or the standard binary model. In the case of the caustic
crossing fits, the data points removed from the lightcurve do not
contribute to $\com$. 

With these classifications in place, we can now define the binary
detection efficiency and the orbital motion detection efficiency. The
binary detection efficiency is the fraction of detectable microlensing
events that show binary signatures
\begin{equation}
\fbs \equiv \frac{\nbs}{\nml},
\label{binaryEfficiency}
\end{equation}
where $\nml$ is the number of events satisfying $\dcbase/\nobs>0.3$,
and $\nbs$ is the number of events satisfying $\dcsingle>200$. The
orbital motion detection efficiency is the fraction of binary events
that show orbital motion signatures
\begin{equation}
\fom \equiv \frac{\nom}{\nbs},
\label{omEfficiency}
\end{equation}
where $\nom$ is the number of events satisfying $\dcstat > 200$.

To be confident of our results we must quantify the effectiveness of
the modelling prescriptions we use. We can do this by measuring the
rate of false positives in our samples. To measure these rates we simulate
both single lens events and static binary lens events, drawn from the
same distributions as the orbiting lens events. These events then go
through the same fitting procedure as the orbiting lens events and
are subject to the same cuts. The binary lens false positive rate
$\fbs^{\rm single}$ is therefore the fraction of detectable single lens
microlensing events that survive the $\dcsingle>200$ cut, and the
orbital motion false positive rate $\fom^{\rm static}$ is the fraction of
static binary lens events that survive the $\dcstat>200$ cut. 

%
%
\section{Results}
\label{Results}

\subsection{What fraction of events show orbital motion?}

\begin{table}
\caption{Summary of the results for planetary lenses.}
\begin{tabular}{@{}lcc}
\hline
Orbit & static & circular \\
\hline
Single & 48511 & 49226 \\
Binary & 1364 & 1366 \\
Caustic & 410 & 449 \\
Caustic static & 397 & 414 \\
Caustic orbital motion & 7 & 35 \\
Smooth & 954 & 917 \\
Smooth static & 931 & 883 \\
Smooth orbital motion & 23 & 34 \\
\hline
\end{tabular}
\label{rawNumbersPlanets}
\end{table}

\begin{table}
\caption{Summary of the results for stellar binary lenses.}
\begin{tabular}{@{}lccc}
\hline
Orbit & static & circular & eccentric\\
\hline
Single & 4151 & 4046 & 4153 \\
Binary & 1413 & 1424 & 1385 \\
Caustic & 641 & 635 & 613 \\
Caustic static & 608 & 538 & 550 \\
Caustic orbital motion & 25 & 86 & 61 \\
Smooth & 772 & 789 & 772 \\
Smooth static & 764 & 743 & 729 \\
Smooth orbital motion & 8 & 46 & 43 \\
\hline
\end{tabular}
\label{rawNumbersStars}
\end{table}

\begin{figure}
\includegraphics[width=76mm]{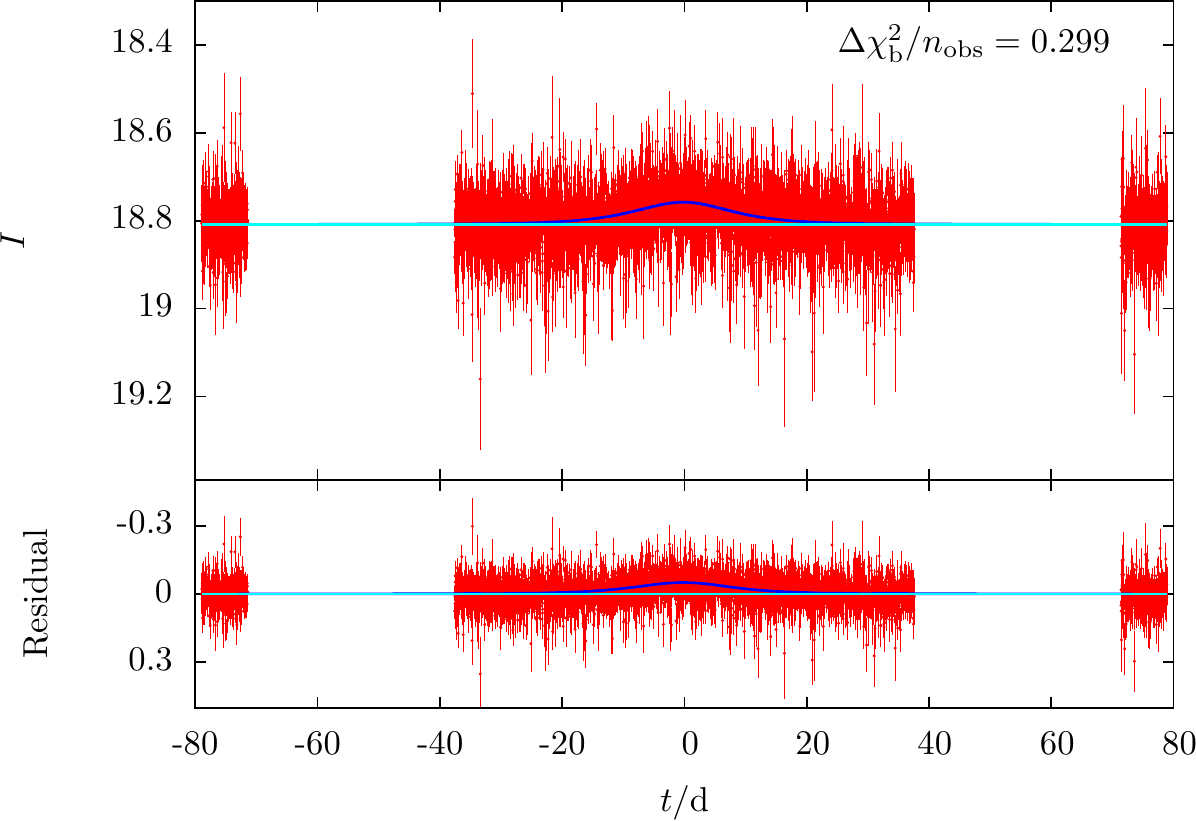}\\
\includegraphics[width=74mm]{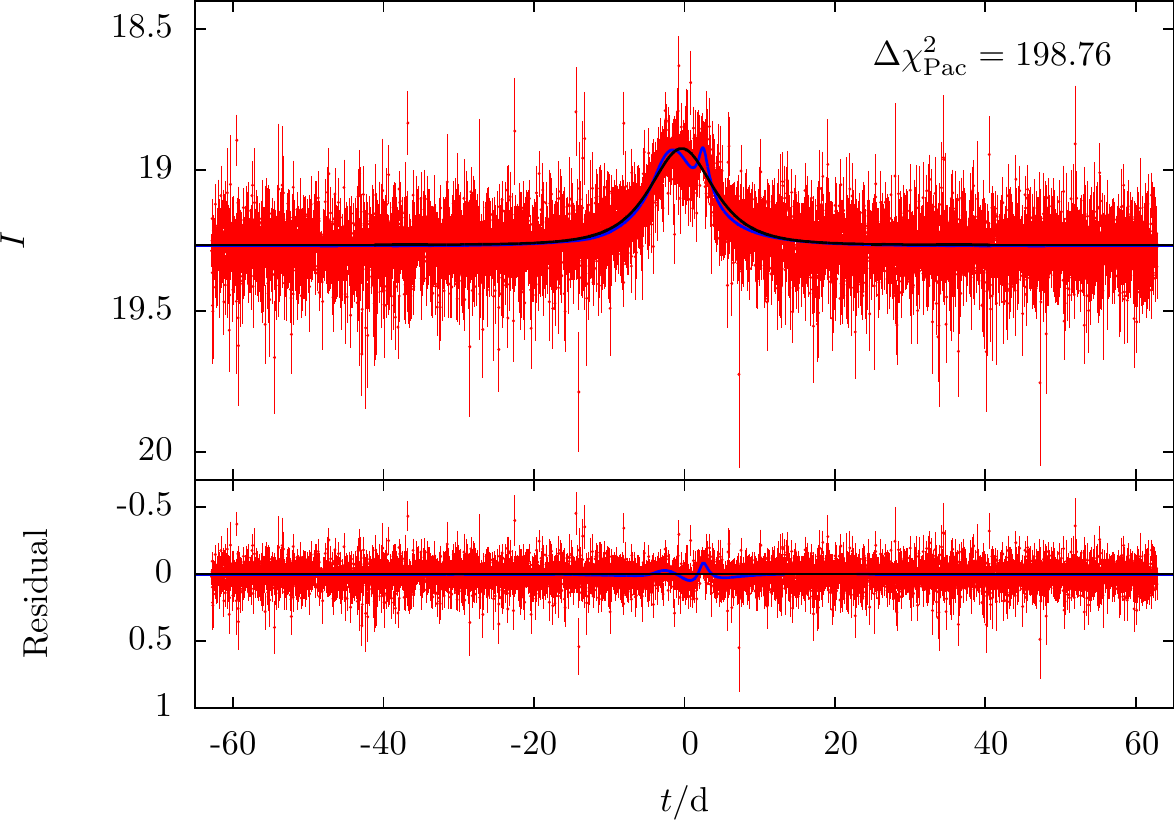}\\
\includegraphics[width=76mm]{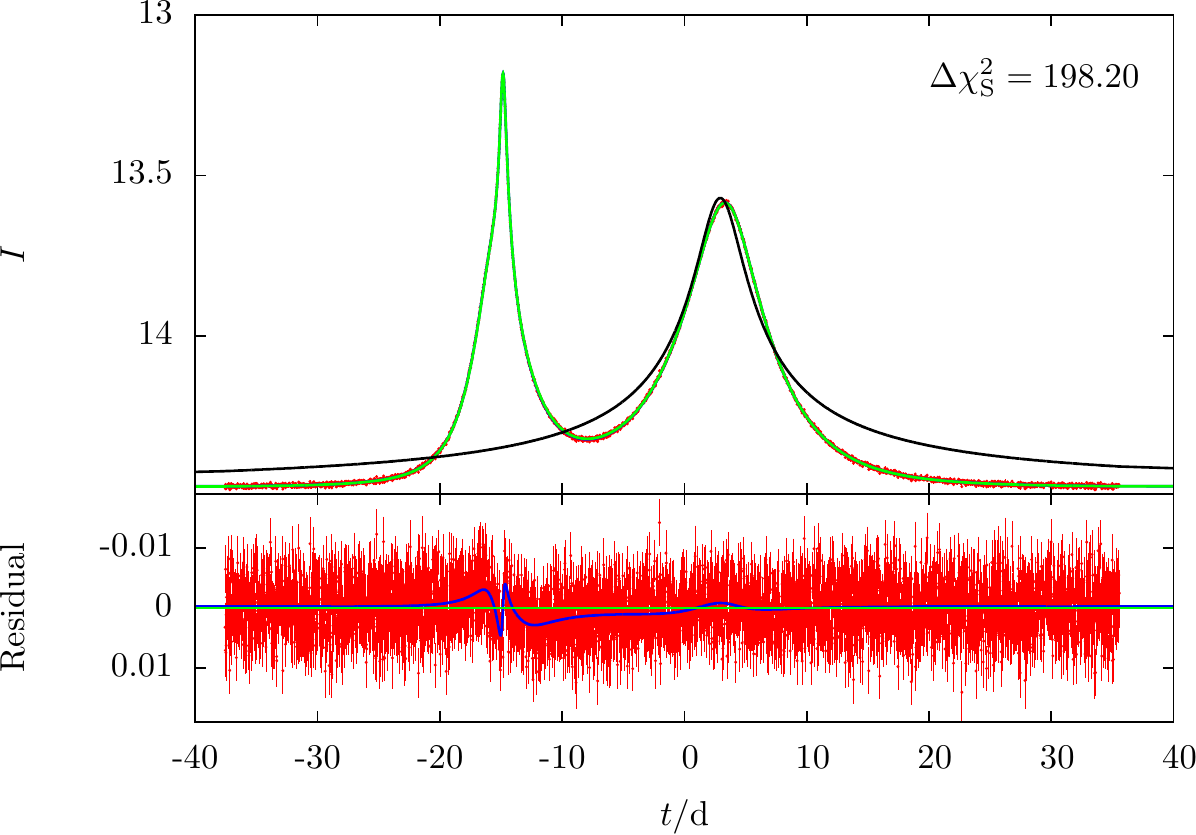}\\
\caption{Example lightcurves of three events that narrowly failed one
  of the classification cuts. From top to bottom, the lightcurves
  failed the $\frac{\dcbase}{\nobs} \equiv \frac{\cbase - \com}{\nobs} >
  0.3$, $\dcsingle \equiv \csingle - \com > 200$, and $\dcstat \equiv
  \cstat - \com > 200$ cuts, respectively. The upper panel of each
  subplot shows the lightcurve, and the lower panel the residual with
  respect to the appropriate model for the cut, i.e. the constant
  baseline `model', the best fitting \paczynski model, and the best
  fitting static binary model in the top, middle and lower subplots,
  respectively. Colour coding is the same as in
  Figure~\ref{lightcurveCoverage}, and the cyan line in the top
  subplot shows the constant baseline `model'. Further details for the
  events can be found in Tables~4 and 5 in the online supplementary
  material.
}
\label{cutExamples}
\end{figure}

We begin by presenting and analyzing the results of the simulations as
a whole, calculating the fraction of microlensing events in which we
expect to see orbital motion events. Tables~\ref{rawNumbersPlanets}
and \ref{rawNumbersStars} summarize the results of the cuts described
in the previous section, for planetary and stellar binary events
respectively. It should be noted that in a small number of caustic
crossing events, the fitting procedure failed, and so these events
have been excluded from the analysis of the orbital motion detection
efficiency, but not of the binary detection efficiency. These events
are included in the Binary and Caustic rows of
Tables~\ref{rawNumbersPlanets} and \ref{rawNumbersStars}, but not in
the others. Figure~\ref{cutExamples} shows some lightcurves which were
slightly below the threshold for each cut. 

\begin{table}
\caption{Binary and orbital motion detection efficiencies.}
\begin{tabular}{@{}lccc}
\hline
Orbit & & circular & eccentric\\
\hline
$q<0.01$ & $\fbs$ & $ 0.0772 \pm 0.0014 $ & --\\
$q<0.01$ Caustic & $\fom$  & $ 0.061 \pm 0.010 $ & --\\
$q<0.01$ Smooth & $\fom$ & $ 0.0130 \pm 0.0055 $ & --\\
$q<0.01$ All & $\fom$  & $ 0.029 \pm 0.005 $ & --\\
\hline
$q\ge 0.01$ & $\fbs$ & $ 0.260 \pm 0.004 $ & $ 0.251 \pm 0.004 $ \\
$q\ge 0.01$ Caustic & $\fom$  & $ 0.098 \pm 0.011 $ & $ 0.060 \pm
0.010 $ \\
$q\ge 0.01$ Smooth & $\fom$  & $ 0.048 \pm 0.006 $ & $ 0.045 \pm 0.006 $ \\
$q\ge 0.01$ All & $\fom$  & $ 0.070 \pm 0.006 $ & $ 0.052 \pm 0.006 $ \\
\hline
\end{tabular}
\label{omeffTable}
\end{table}

Table~\ref{omeffTable} shows the binary detection efficiency and
orbital motion detection efficiency for both planetary and stellar
binary lenses. It should be noted that the binary detection efficiency
will be larger than for microlensing events with finite sources, as
the effect of the finite source will be to smooth out sharper
lightcurve features, and usually reduce the amplitude of deviations
from the single lens model. This means that $\fbs$ for planetary
lenses is likely a significant overestimate, however, for stellar
binary lenses the result is likely to be more realistic as binary
lightcurve features tend to be stronger and have longer durations. The
detection efficiencies presented have been corrected for systematic
false positives from each fitting stage by subtracting the measured
false positive rates $\fbs^{\rm single}$ and $\fom^{\rm static}$ from
the detection efficiencies measured for orbiting lenses. From a
simulation of $10^4$ single lenses with no false positives we measured
$\fbs^{\rm single} = 0_{-0}^{+4.7\times 10^{-5}}$, where the error
quoted is a statistical $1\sigma$ confidence limit calculated using
Wilson's score method~\citep{Wilson:1927fcl,Newcombe:1998acl}. To
calculate the errors on the corrected detection efficiencies shown in
the table, and on those we present in the next subsection, we use
Wilson's score method adapted for the difference of two
proportions~\citep[method 10]{Newcombe:1998dcl}. For planetary events
we measured false positive rates of $\fom^{\rm static} =
\ase{0.0241}{0.0032}{0.0036}$ for smooth events and $\fom^{\rm static}
= \ase{0.0173}{0.0039}{0.0050}$ for caustic crossing events. For
stellar binary events we measured $\fom^{\rm static} =
\ase{0.0104}{0.0022}{0.0028}$ for smooth events and $\fom^{\rm static}
= \ase{0.0395}{0.0050}{0.0056}$ for caustic crossing events. The
overall orbital motion detection efficiencies were calculated as a
weighted average of the detection efficiencies for smooth and caustic
crossing events, once corrected for false positives.

While in many cases we may not be able to say that a lightcurve in our simulations definitively shows orbital motion signatures, due to relatively high rates of false positive detections, there is a clear excess of detections in the circular and eccentric orbit simulations relative to the static ones, though the detection of this excess is only marginal in smooth planetary events. Interestingly, there appears to be a discrepancy in the orbital motion detection efficiencies for stellar binary caustic crossing events. The same static orbit simulation results were used to calculate the corrected orbital motion efficiencies for both circular and eccentric orbits, which means that the measurements are not independent. Also, the eccentricity of the orbits allows the projected separation to take a wider range of values than the circular orbits, which means the false positive rate measured with the same distribution for circular orbits is likely an overestimate for eccentric orbits; for caustic crossing events the majority of false positives are caused by events with resonant caustic topology (see Figure~\ref{d0_q-scatter} later in this section). We therefore believe the discrepancy to be caused largely due to a combination of a relatively large statistical fluctuation in the number of eccentric orbit events that do show orbital motion, and an overestimate of the false positive rate for eccentric orbits. 

\subsection{What affects the detectability of orbital motion?}

We now investigate the effects that various system parameters have on the detectability of orbital motion. We look at the dependence of the orbital motion detectability on both the standard microlensing parameters and the physical orbital parameters, and compare them where appropriate. While we conducted two sets of simulations, one with circular orbits and one with eccentric orbits, we only present the results for those with circular orbits here, as both sets are in good agreement.

%
%
\begin{figure}
\includegraphics[width=84mm]{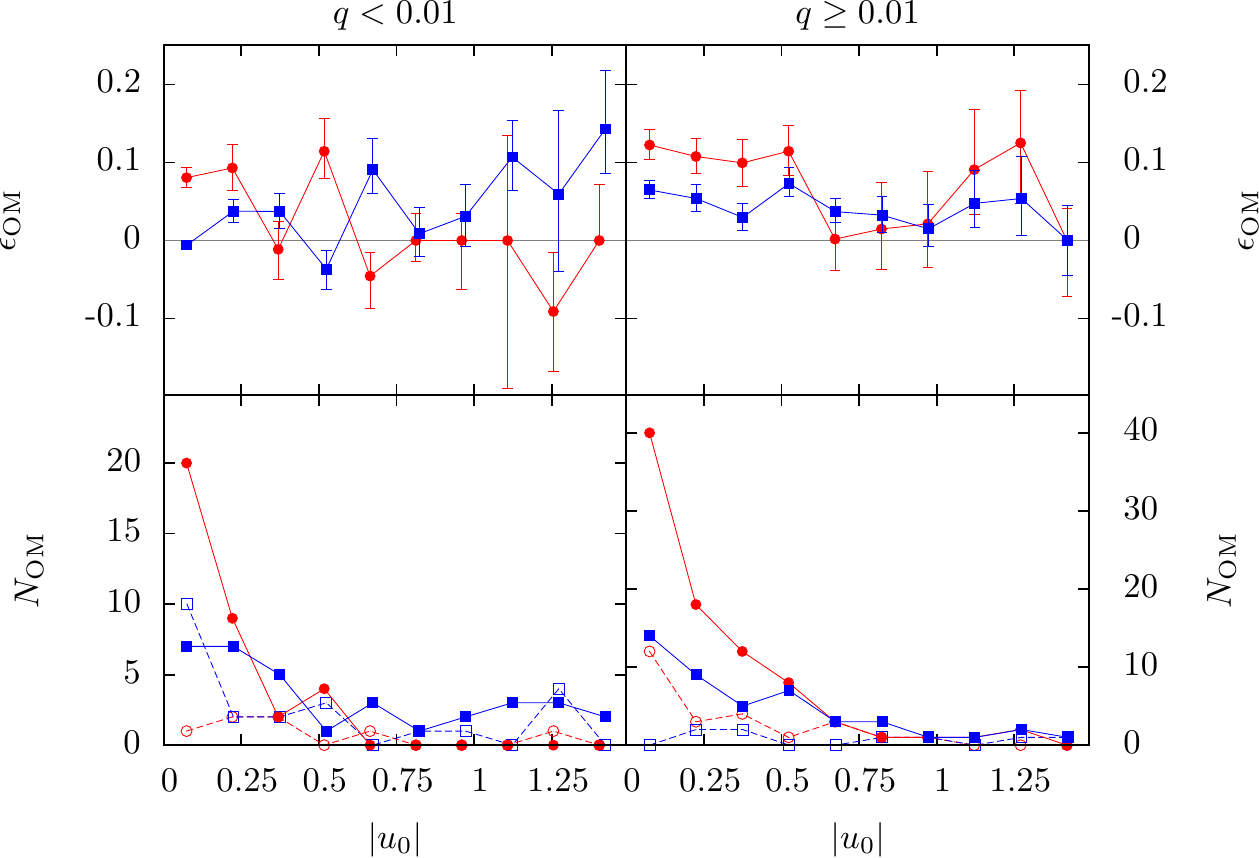}
\caption{Plot of the orbital motion detection efficiency, corrected
  for systematic false positives (top panels), and the absolute number
  of orbital motion detections in the simulations (lower panels),
  against the impact parameter $|\uzero|$. Results are shown for
  lenses with planetary mass ratios (left) and binary star mass ratios
  (right). Red lines with filled squares show the results for caustic
  crossing events and blue lines with filled circles show the results
  for smooth events. In the upper panels a line marks zero orbital
  motion detection efficiency. All events had circular orbits, and in
  the lower panels results are shown for events where the lens
  components were in orbit (solid lines, filled points) and where they
  were held static for the calculation of the false positive rate
  (dashed lines, open points). Events have been binned into bins of
  equal width, and points plotted at the centre of the bin. Note that
  in the lower panels the scales are different.}
\label{u0-OM}
\end{figure}

We begin by looking at the dependence on the impact parameter, the
sole parameter that determines the maximum magnification of a single
lens microlensing event
$A_{\mathrm{max}}=(\uzero^2+2)/(\uzero\sqrt{\uzero^2+4})$~\citep{Paczynski:1986gml}.
For all binary lenses, except wide stellar binaries, the central
caustic is located near to the centre of mass, and so $\uzero$
determines whether or not the source will encounter this caustic. Figure~\ref{u0-OM} plots the orbital motion detection efficiency as a fraction of caustic crossing or smooth binary events (top panels), and the total number of orbital motion detections (bottom panels), against the impact parameter for both planets (left panels) and binary stars (right panels). In the plots, red lines represent data for caustic crossing events and blue lines for smooth events. In the top panel the orbital motion detection efficiency has been corrected for systematic false detections as described in the previous subsection, whereas the bottom panel shows the number of detections for both orbiting (solid lines, filled points) and static lenses (dashed lines, open points). Note that the orbital motion detection efficiency can be negative, due to statistical fluctuations, but if it is, the measurement should be consistent with zero. The events have been binned into bins of constant width, on the scale that they are plotted. It should also be noted, that the number of planetary events simulated was a factor of 9 larger than the number of stellar binary events.

The plots of orbital motion detection efficiency (from here on
detection efficiency) against $|\uzero|$ for caustic crossing events
show much the same trends for both planetary and stellar binary
lenses, with significant detection efficiencies for high-magnification
(low $|\uzero|$) events only, with no caustic crossing planetary
detections for $|\uzero|\gtrsim 0.6$, and only a few for stellar
binaries. This is due to the location of central and resonant caustics
close to the center of mass, which the source can only cross in events
with small $|\uzero|$. Consequently, for the events with larger
$\uzero$, the source can only cross weaker secondary caustics, which
in the case of wide binaries will typically move slowly, and in the
case of close binaries are typically very small and are rarely
crossed. The secondary caustics of close stellar binaries are
significantly larger and stronger than those of planetary lenses, and
so the chances of the source crossing them is higher, and the caustic
has a longer time in which to change due to orbital motion as the
source crosses it, leading to the small positive efficiency for
$|\uzero|\gtrsim 0.6$. For smooth events, the planetary and stellar
binary lenses show weak but opposing trends, with the efficiency
increasing slightly as $|\uzero|$ increases for planetary events and
decreasing slightly as $|\uzero|$ increases for stellar binary events,
indicating that the impact parameter only plays a small role in
orbital motion detectability for smooth lightcurves. Note however,
that for both smooth and caustic crossing events the number of orbital
motion detections, as opposed to the detection efficiency, is a strong
function of $|\uzero|$, peaking at small values, due to the dependence
of the binary detection efficiency on the impact parameter.

%
%
\begin{figure}
\includegraphics[width=84mm]{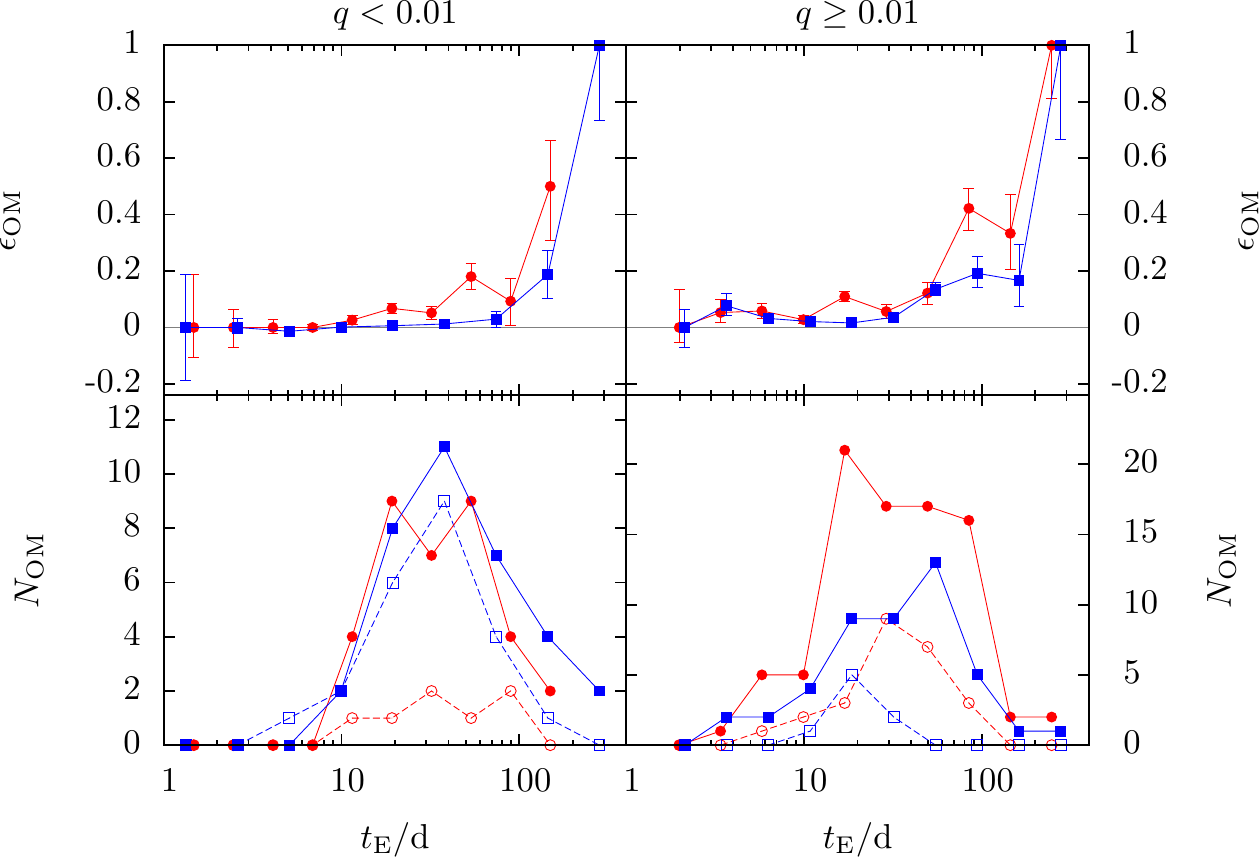}
\caption{As Figure~\ref{u0-OM}, but plotted against the event time-scale $\tein$.}
\label{tE-OM}
\end{figure}

Figure~\ref{tE-OM} plots the detection efficiency against the event time-scale $\tein$. All classes of binary event (planetary or binary, smooth or caustic crossing) show a strong detection efficiency dependence on the event time-scale. The reason for this dependence is simply because a longer time-scale allows the lens to complete a larger fraction of its orbit, and hence cause a larger change in the magnification pattern, during the time in which the source probes regions of the magnification pattern that deviate from that of a single lens. In the case of planetary lenses, it seems that a time-scale of greater than $\sim 10$~days is necessary for caustic crossing events and slightly longer for smooth events. Caustic crossing events show larger detection efficiency than smooth events, even at shorter time-scales. This is likely due to the high accuracy with which caustic crossing times, and the lightcurve shape around caustic crossings can be measured. In the case of OGLE-2006-BLG-109 this has allowed the orbital motion of the lens to be measured from data covering just $\sim 0.2$~percent of the orbit~\citep{Gaudi:2008jsa,Bennett:2010jsa}. Smooth events in contrast require a much larger fraction of the orbit to cause significantly detectable changes in the lightcurve, and hence require a longer time-scale to achieve the same detection efficiency. However, typically it is possible for smooth features to cover a much larger fraction of the lightcurve than caustic crossing features, lessening the effect of this discrepancy.

For stellar binary lenses, orbital motion features can be can be detected effectively over almost the entire range of time-scales that we simulated, though with a low efficiency for time-scales below $\sim 40$ days for smooth events and $\sim 10$ days for caustic crossing events. For events with time-scales over $\sim 100$~days, the detection efficiency reaches $\sim 20$~percent for smooth events and $\sim 40$~percent for caustic crossing events. The detection efficiencies are similar for planetary events. The majority of planetary and binary events showing orbital motion have time-scales of around $\sim 10$--$40$~days, with few events at larger $\tein$ due to the steep $\tein^{-3}$ distribution at large time-scales~\citep{Mao:1996mm}. However, the strong dependence of $\fom$ on time-scale means that the slope of the high $\tein$ tail of the distribution of orbital motion events is much shallower than $\tein^{-3}$.

%
%
\begin{figure}
\includegraphics[width=84mm]{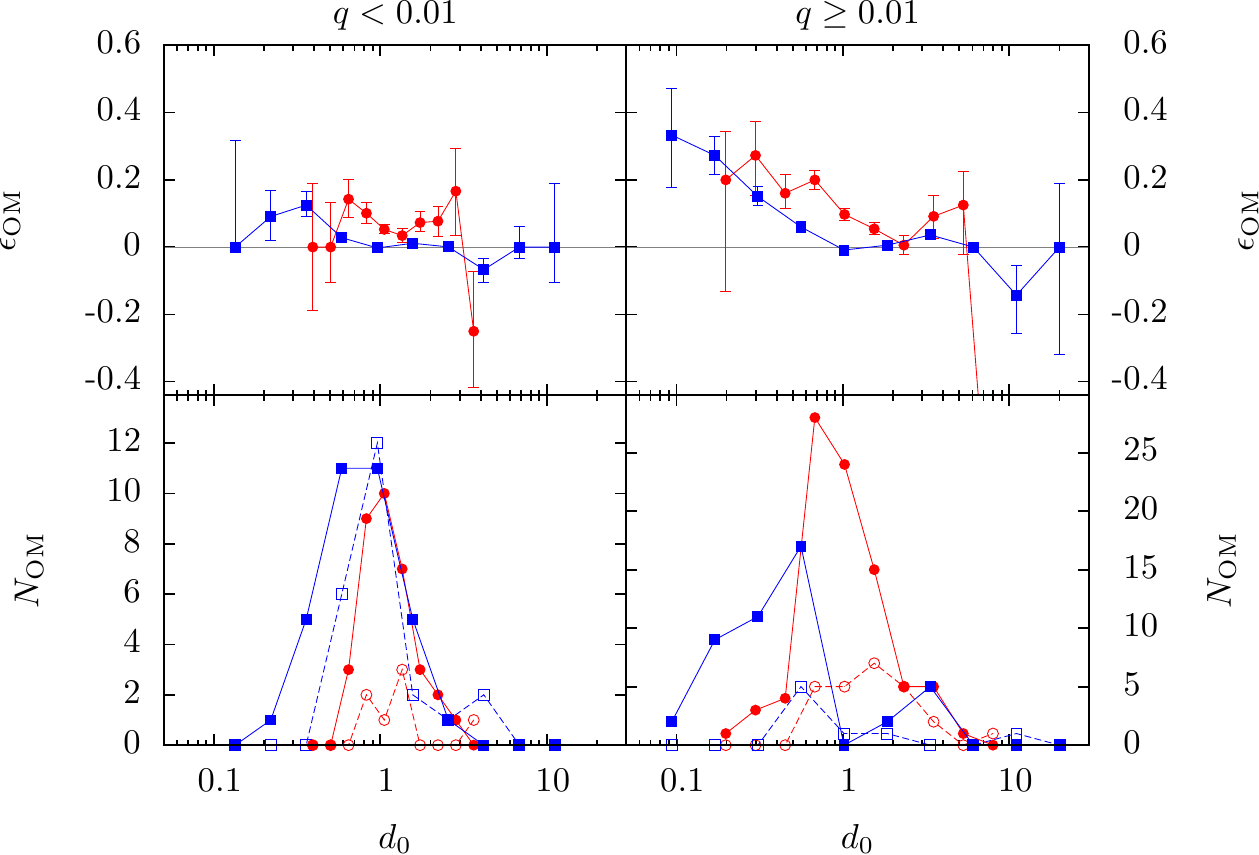}
\caption{As Figure~\ref{u0-OM}, but plotted against $\dzero$ the lens separation at time $\tzero$.}
\label{d0-OM}
\end{figure}

%
%
\begin{figure}
\includegraphics[width=84mm]{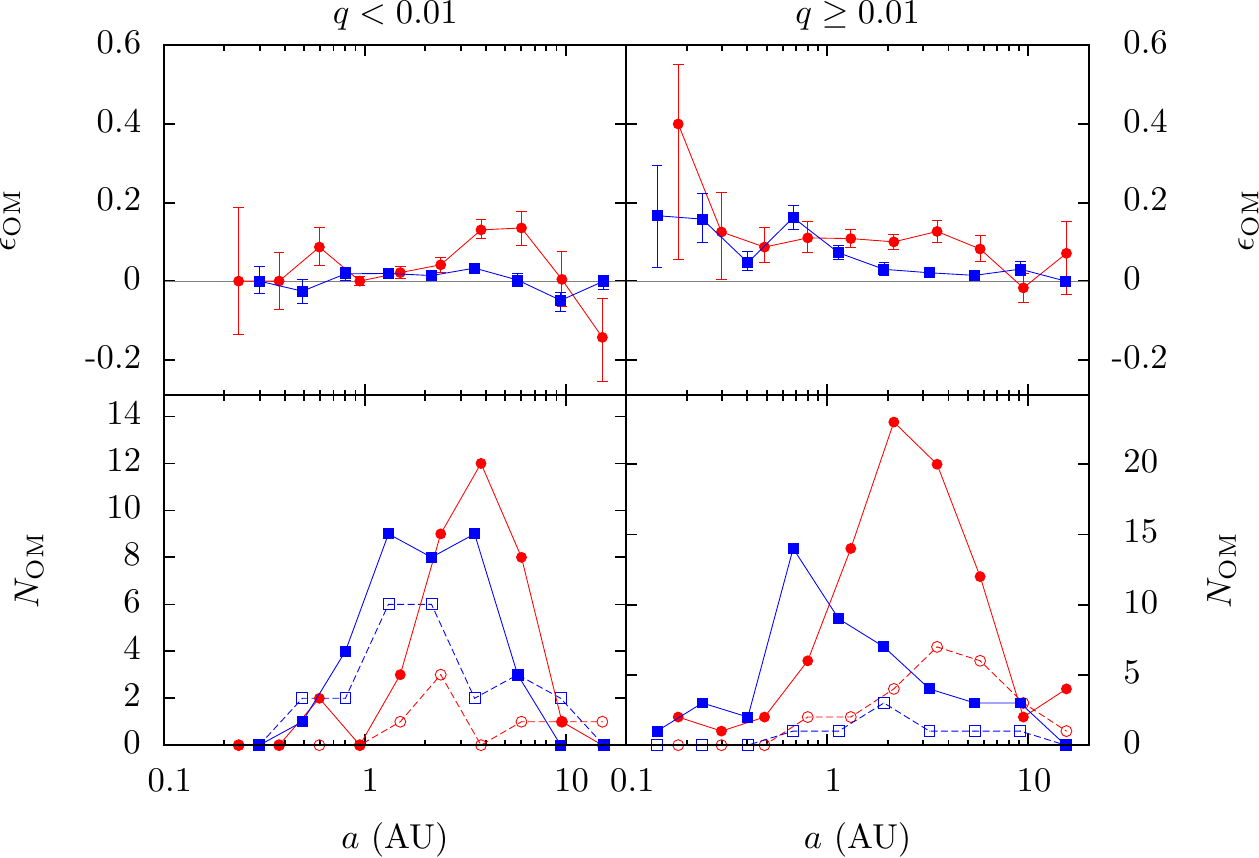}
\caption{As Figure~\ref{u0-OM}, but plotted against the semimajor axis $a$.}
\label{a-OM}
\end{figure}

The plots of detection efficiency against projected separation
$\dzero$ (Figure~\ref{d0-OM}) and semimajor axis $a$
(Figure~\ref{a-OM}) tell largely the same story. The detection
efficiency in stellar binaries has a significant inverse dependence on
both $\dzero$ and $a$, as would be expected from the dependence of
orbital velocity of semimajor axis. However, the behaviour for
planetary lenses is less intuitive: for caustic crossing events, there
is a significant peak in the detection efficiency at $a \sim 4$~AU,
and a peak/shoulder at $\dzero \sim 2$. There is a second peak in
$\fom$ with $\dzero$. The two peaks occur at values of $\dzero$ where
the boundaries between caustic topologies occur for the highest mass
ratio planets. It is at these boundaries that, for a small change in
projected separation $\dd(\log d)$, the largest changes in the
caustics occur. The peak in $\fom$ against $a$ at $a\sim 4$~AU for
caustic crossing planetary events is accompanied by a hint of a peak
at small values of $a$. The peak at $a\sim 4$~AU can be explained by considering the typical scale of the Einstein ring, and by considering the trend of $\fom$ with the event time-scale. The typical size of the Einstein ring for a microlensing event is $2$-$3$~AU, but as seen in Figure~\ref{tE-OM}, orbital motion effects typically occur in events with larger time-scales. As the time-scale is correlated with the Einstein ring size, and caustic crossing events typically occur in systems with $\dzero \sim 1$, the peak orbital motion detection efficiency occurs at a semimajor axis slightly above the typical Einstein ring size, at $a\sim 4$~AU. The increase in orbital velocity as $a$ decreases likely causes the second weaker peak in $\fom$ at smaller $a$. Little can be said about the trend of $\fom$ with $a$ for smooth planetary events as small numbers of events, and the distribution of Einstein radius sizes serves to smear out any obvious trends. However, when plotted against $\dzero$, $\fom$ does increase towards smaller values of $\dzero$, as would be expected from orbital velocity considerations.

Returning to the caustic crossing stellar binary events, $\fom$
flattens off as $a$ increases to $\sim 4$~AU, before dropping to
zero. This flattening likely has the same cause as the peak for
planetary caustic crossing events. We see the more intuitive inverse
trend in stellar binaries because of the stronger and larger
magnification pattern features that they exhibit, and the larger range
of $d$ over which the caustics have a significant size. This results
in a distribution of events over $a$ and $\dzero$ which is broader and
somewhat less peaked than for planetary events (see the lower panels
of Figures~\ref{d0-OM} and \ref{a-OM}). This allows the inverse
relationship between orbital velocity and semimajor axis to have a
greater influence on the trend in the orbital motion detection efficiency. We note that the
reason we see such a complicated relationship between $\fom$ and $a$
and $\dzero$, but not for example between $\fom$ and $\tein$, is that
the orbital separation affects the orbital velocity in a relatively
simple way and caustic size and strength in a complicated way, whereas
$\tein$ only affects, or more accurately is the result of, a fairly
simple dependence on a single factor in the detection of orbital
motion, the source speed.

%
%
\begin{figure}
\includegraphics[width=84mm]{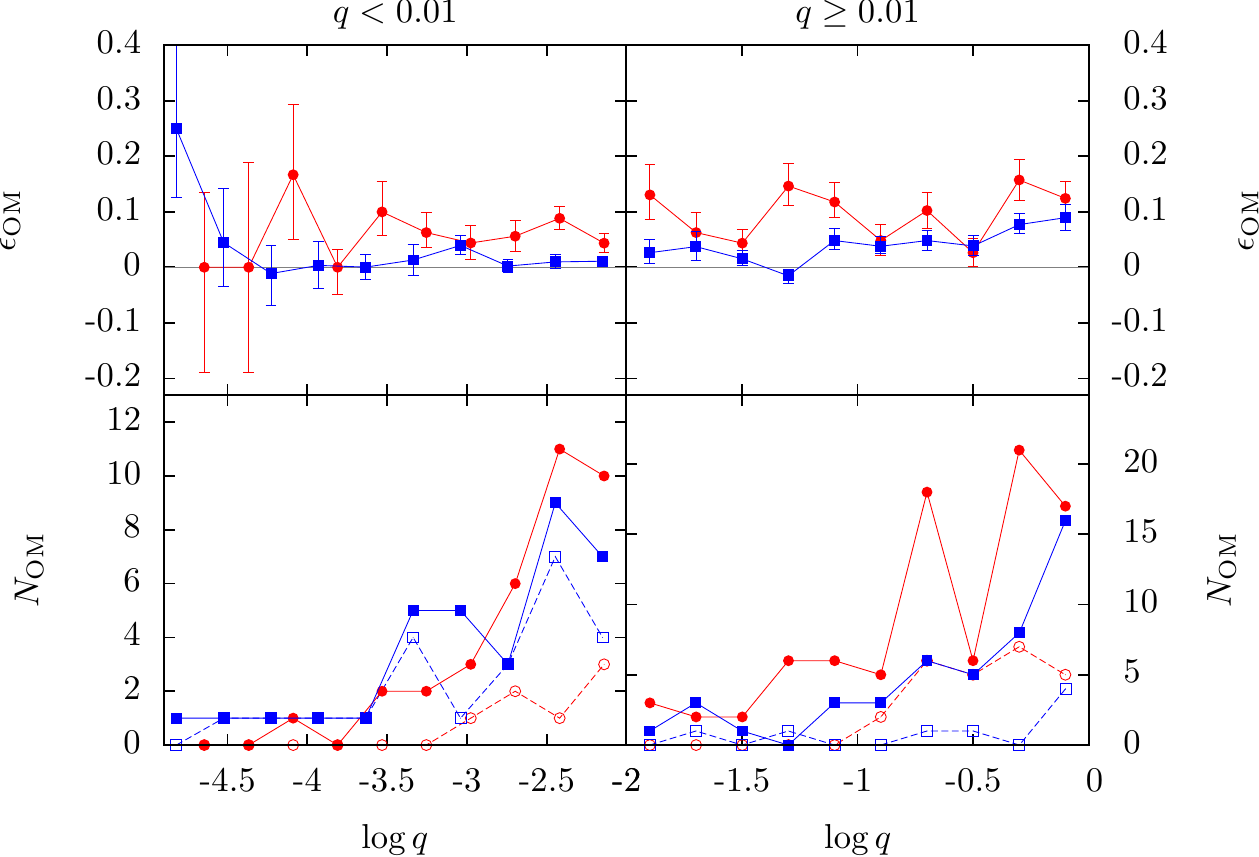}
\caption{As Figure~\ref{u0-OM}, but plotted against the mass ratio $q$.}
\label{q-OM}
\end{figure}

Figure~\ref{q-OM} plots the detection efficiency against the mass
ratio $q$. Treating both planetary and stellar binary lenses together,
there is a trend of increasing detection efficiency with increasing
$q$, for both smooth and caustic crossing events. However, for caustic
crossing events, this increase is very shallow, with a factor of
$\lesssim 3$ increase over three decades in $q$, from $\log q \approx
-3$ to $\log q = 0$. For smooth events, there is a stronger trend,
with the detection efficiency being effectively zero for $\log
q\lesssim -3.5$, while rising from $\sim 1$~percent to $\sim
10$~percent over the range $-3.5 \lesssim \log q < 0$. These shallow
dependencies are somewhat unexpected in relation to the somewhat
stronger $q^{0.5}$ dependence of the binary detection efficiency,
which derives directly from the dependence of caustic size on
$q$~\citep{Han:2006pcp}. However, the orbital detection efficiency effectively divides
through by this dependence (unlike the curves of the number of orbital
motion detections, which show a strong dependence on $q$), to leave a
very shallow orbital motion detection efficiency curves. The other
effect that $q$ has on the lightcurve features is to make them
stronger as $q$ increases. In caustic crossing events the caustic
features are usually strong, independent of the value of $q$, and
hence the caustic crossing events curve is shallower than the curve
for smooth events, for which the dependence of the feature strength on
$q$ is much more important. 

%
%
\begin{figure}
\includegraphics[width=84mm]{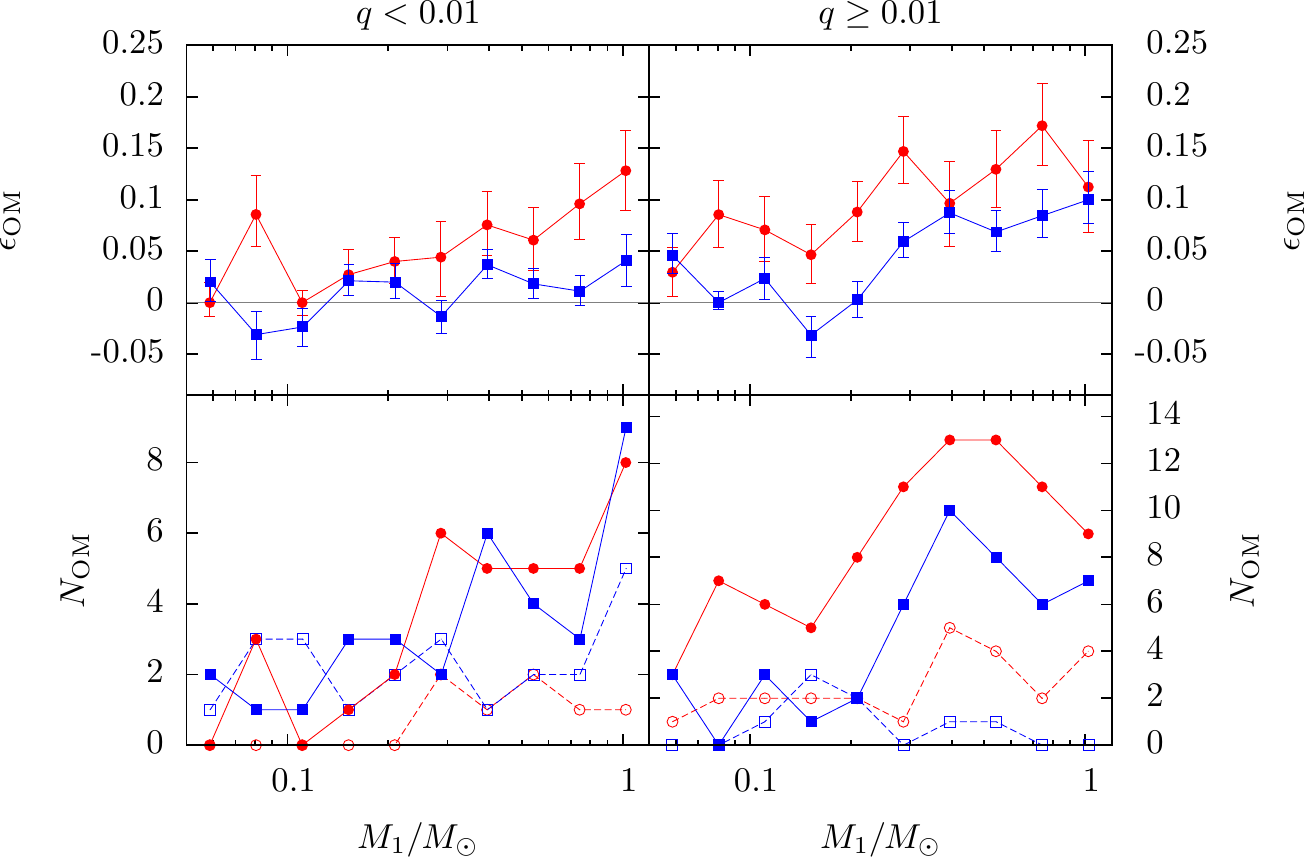}
\caption{As Figure~\ref{u0-OM}, but plotted against the primary lens mass $\bigmone$.}
\label{M1-OM}
\end{figure}

Figure~\ref{M1-OM} shows the detection efficiency plotted against the primary lens mass. The dependence is as expected for both mass ratio regimes and for both types of binary event, increasing as the mass of the primary increases. The trend is strongest in smooth, stellar binary events.

%
%
\begin{figure}
\includegraphics[width=84mm]{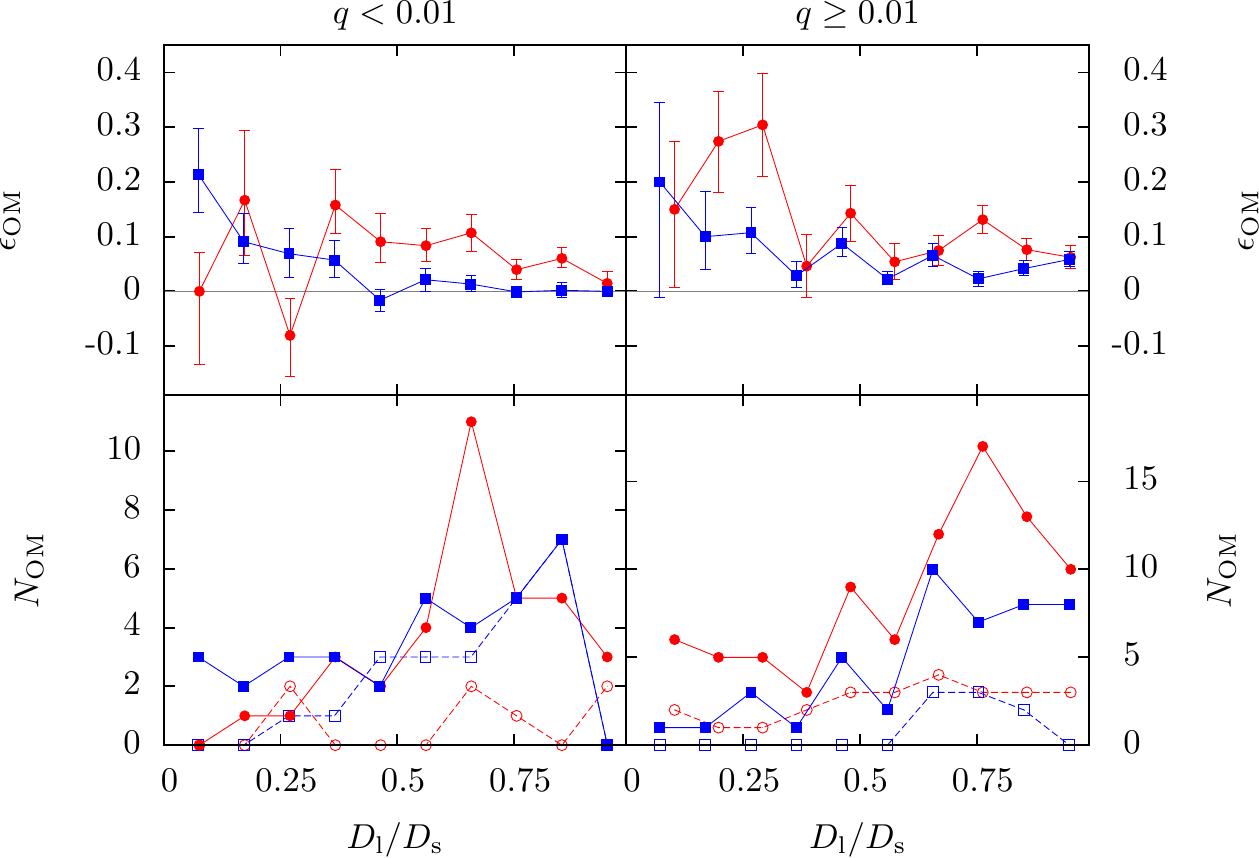}
\caption{As Figure~\ref{u0-OM}, but plotted against the lens distance $\dl$.}
\label{x-OM}
\end{figure}

Figure~\ref{x-OM} plots the detection efficiency against the lens
distance. In all cases, a trend of increasing detection efficiency
with decreasing lens distance is seen, though caustic crossing events
suffer from small number statistics at low values of $\dl/\ds$. Note
however, that the frequency distribution (plotted in the lower panels of
Figure~\ref{x-OM}) of orbital motion events, once false positives have
been approximately accounted for, is different, being peaked at
$\dl/\ds \sim 0.7$.

%
%
\begin{figure}
\includegraphics[width=84mm]{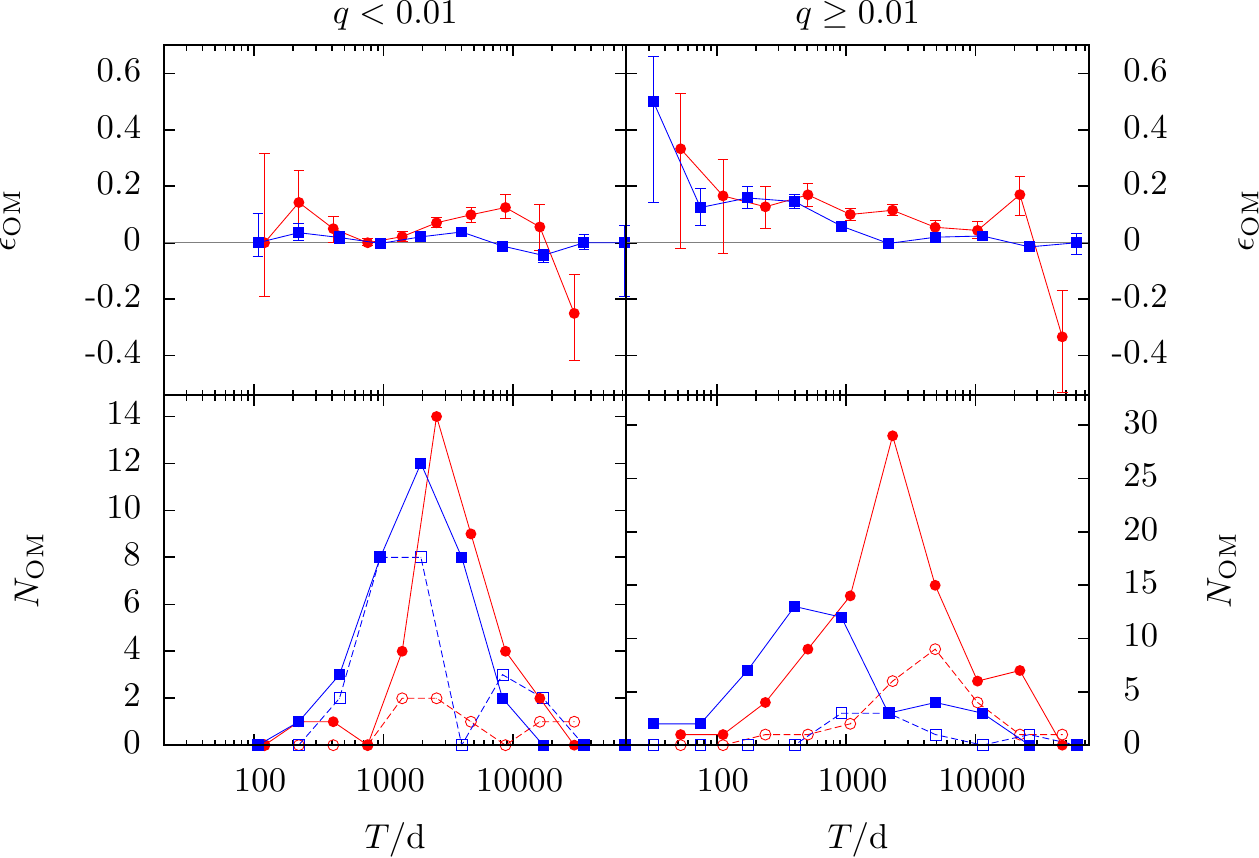}
\caption{As Figure~\ref{u0-OM}, but plotted against the orbital period $T$.}
\label{T-OM}
\end{figure}

Figure~\ref{T-OM} shows the detection efficiency plotted against the
orbital period. Both types of stellar binary event show a significant
inverse trend. Planetary caustic crossing events show a peak, and
stellar caustic crossing events a flattening, at large periods. These
features correspond directly to similar features in the curves of
$\fom$ with $a$ and will have the same cause.

%
%
\begin{figure}
\includegraphics[width=84mm]{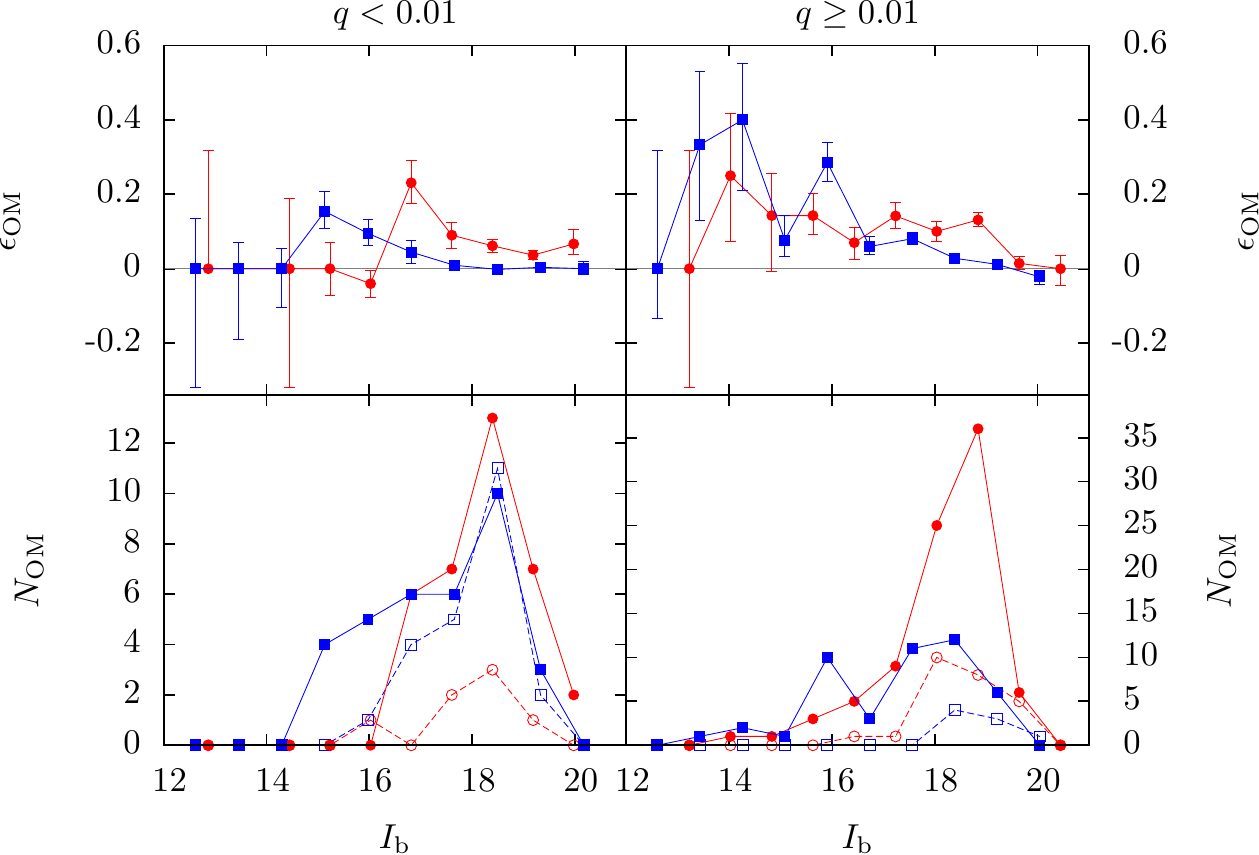}
\caption{As Figure~\ref{u0-OM}, but plotted against the baseline magnitude $\mzero$.}
\label{I0-OM}
\end{figure}

%
%
\begin{figure}
\includegraphics[width=84mm]{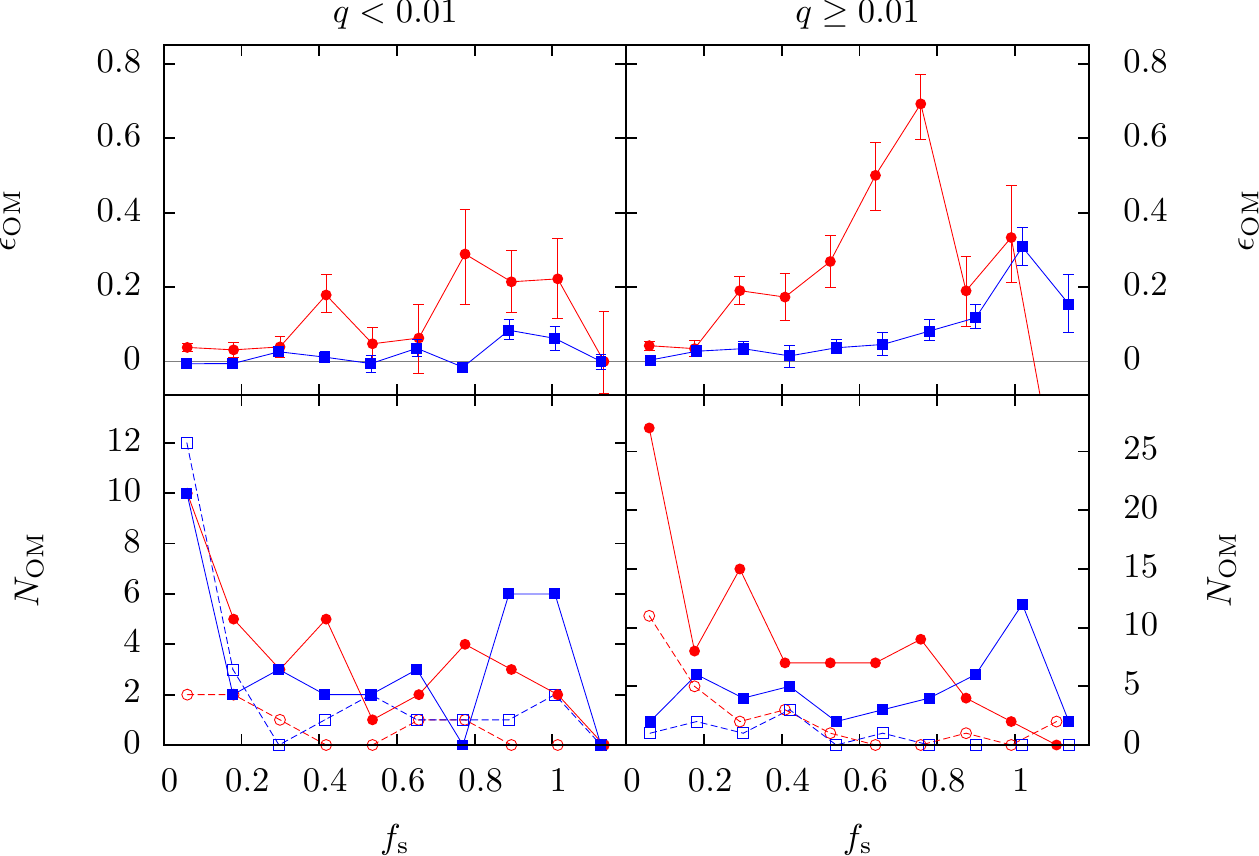}
\caption{As Figure~\ref{u0-OM}, but plotted against the fraction of baseline flux associated with the source $\blendfs$.}
\label{fs-OM}
\end{figure}

Figures~\ref{I0-OM} and \ref{fs-OM} plot the detection efficiency
against the baseline magnitude $\mzero$ and blending fraction
$\blendfs$ respectively. For our purposes, the primary effect of both
parameters is to affect the accuracy with which microlensing
variations can be measured in the lightcurve. For a fixed observing
setup, the baseline magnitude determines the photometric accuracy,
which should lead to a trend of increasing detection efficiency with
decreasing magnitude. This is seen to a certain extent in all cases,
but brighter events may suffer significantly from blending, due to
faint source stars falling entirely within the large point spread
function of a much brighter star. Blending determines the relative
strength of features in the lightcurve, and as such has a much more
significant effect on the detection of smooth binary features, which
have a continuous range of shapes and sizes, compared to the effect on
caustic crossings which are typically sharp and very strong, at least
when finite sources are not considered. It is no surprise, therefore,
that smooth stellar binary events show a significant increase in
orbital motion detection efficiency with blending fraction. This is
less obvious in planetary lenses, likely because the smooth lightcurve
features of planetary lenses are often very weak and difficult to
detect even without the hindrance of the blending, and would not
permit the measurement of higher order effects for any value of
blending fraction. It is more surprising, perhaps, that caustic
crossing events show a significant dependence on blending, as in the
simulations all caustic crossing events were detected as binaries,
regardless of blending. This implies that, at least in some orbital
motion detections in caustic crossing events, the additional smooth
features in the lightcurve, such as peaks and shoulders due to cusp
approaches outside the caustic, and features due to fold caustic
approaches within the caustic, play an important role in the detection
of orbital motion (e.g. lightcurves a and e in
Figure~\ref{separationalExamples} in the next subsection).

%
%
\begin{figure}
\includegraphics[width=84mm]{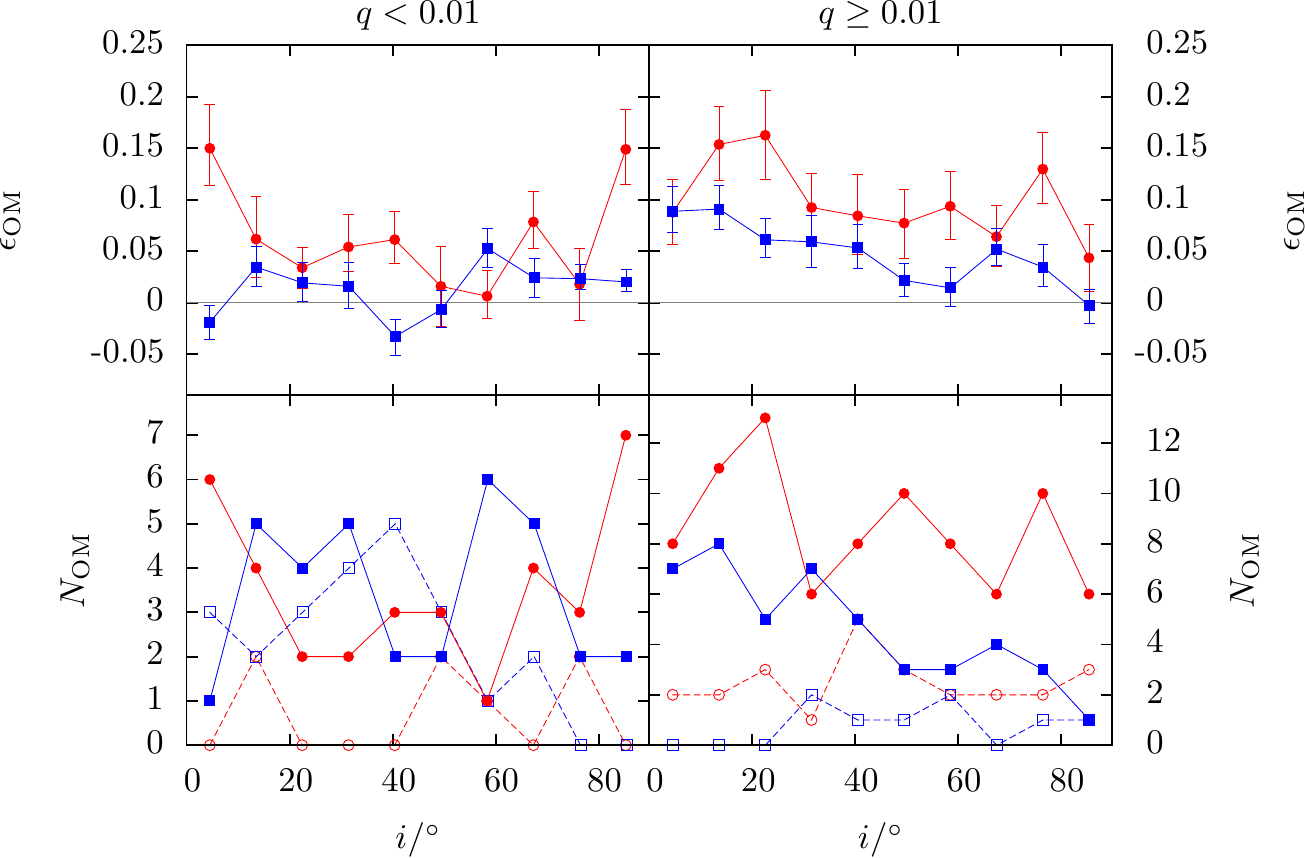}
\caption{As Figure~\ref{u0-OM}, but plotted against the orbital inclination $i$}
\label{i-OM}
\end{figure}

Figure~\ref{i-OM} plots the detection efficiency against
inclination. There is little evidence for any significant dependence
on inclination for all caustic crossing events, and for smooth
planetary events. There is however a stronger trend for smooth stellar
binary events, the detection efficiency decreasing as the inclination
increases. This would be expected in systems where $a/\re \lesssim
\dc$, the boundary between close and resonant caustic topologies, where a reduction in the projected separation due to inclination
would reduce the size of the caustics and reduce the detectability of
both binary features and orbital motion signatures. Unfortunately, due
to the similar effects of inclination and eccentricity on the
projected orbit, the data from the eccentric orbit simulations did not
show any dependence of $\fom$ with eccentricity. This however implies
that the effects of eccentricity on the orbital motion detection
efficiency are not likely to be significantly stronger than those of
inclination.

%
%
\begin{figure}
\includegraphics[width=84mm]{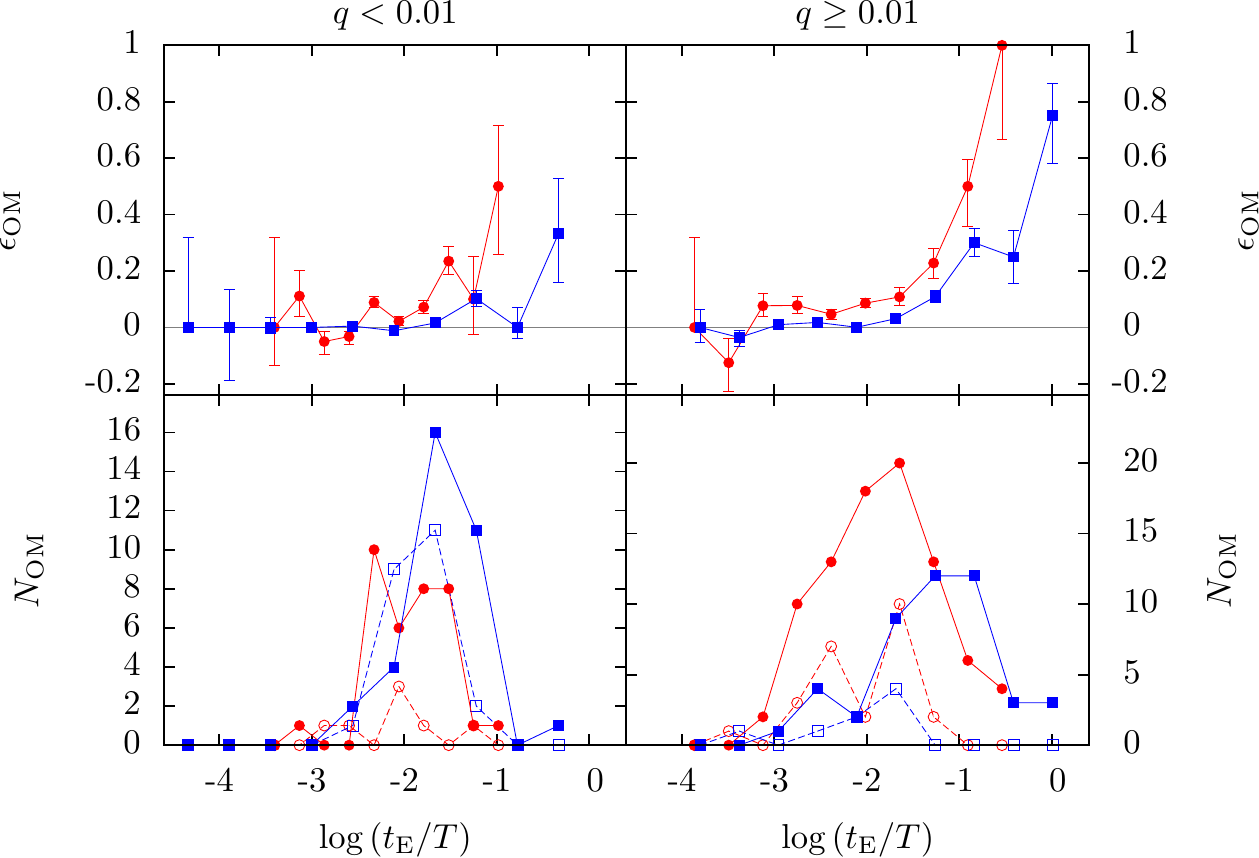}
\caption{As Figure~\ref{u0-OM}, but plotted against the ratio of microlensing to orbital time-scales $\rt=\tein/T$}
\label{tET-OM}
\end{figure}

%
%
\begin{figure}
\includegraphics[width=84mm]{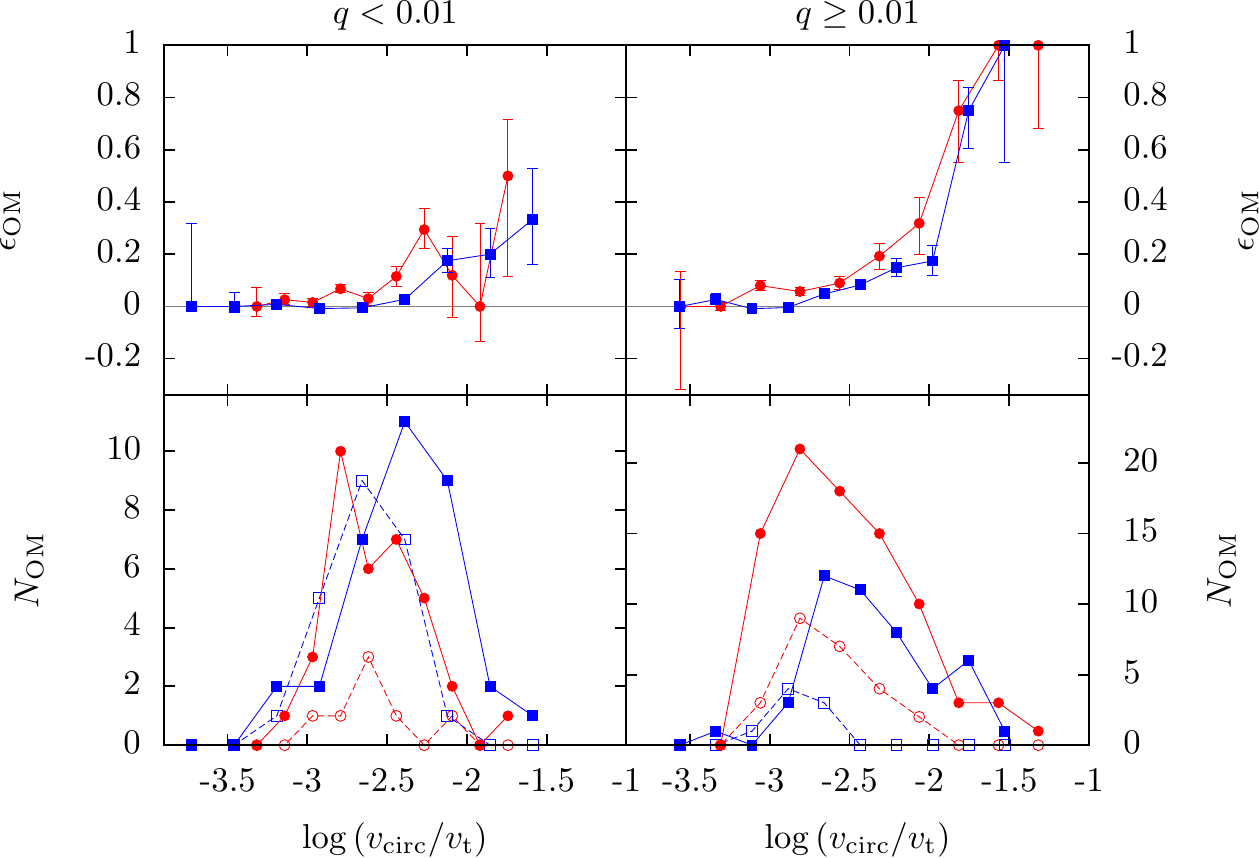}
\caption{As Figure~\ref{u0-OM}, but plotted against the ratio of orbital and source velocities $\rv=v_{\mathrm{circ}}/\vt$}
\label{vcvt-OM}
\end{figure}

It is important not just to consider the system parameters in isolation, but also their combined effects on the orbital motion detection efficiency. For example, \citet{Dominik:1998mrb} introduced two dimensionless ratios to describe the magnitude of orbital motion effects on a binary lens:
\begin{equation}
\rt = \frac{\tein}{T},
\label{timescaleRatio}
\end{equation}
the ratio of time-scales, and
\begin{equation}
\rv = \frac{\vcirc}{\vt},
\label{velocityRatio}
\end{equation}
the ratio of velocities, where $\vcirc=a/2\pi T$ is the circular
velocity of the orbit. These ratios attempt to encapsulate the most
important factors that determine if an event will show orbital motion
features. Figures~\ref{tET-OM} and \ref{vcvt-OM} plot the detection
efficiency against $\rt$ and $\rv$ respectively. Both ratios prove to
be good descriptors of the orbital motion detection efficiency, with
$\fom$ showing strong increasing trends as $\rt$ and $\rv$ increase,
across all mass ratios and lightcurve types, though with a lower
significance in planetary events. It would even seem that, in the case
of smooth events, there exists a threshold value of the ratios, below
which the orbital motion detection efficiency is negligible. For the
ratio of time-scales, the threshold is $\log\rt\approx -2$ for both
planetary and stellar binary lenses, while for the ratio of velocities
the value appears to be more dependent on the mass ratio, taking
values of $\log\rv \approx -2.5$ for planetary lenses, and $\log\rv
\approx -2.75$ for stellar binary lenses. There may be similar
thresholds for caustic crossing events, but at smaller values of $\rt$
and $\rv$.

\subsection{Are there two classes of orbital motion event?}

\citet{Gaudi:2009pmc} has suggested that orbital motion can affect the
lightcurves of microlensing events in two ways. In the first scenario,
the orbital motion effects are dominated by rotation in the lens, as the
orientation of binary axis changes during the time between two widely
separated lightcurve features. The second type of effect is due to
changes in the projected separation over the course of a single
lightcurve feature such as a resonant caustic crossing. In this
subsection we will describe the typical features of each type of event
before investigating to what extent orbital motion events can be
classified in such a way.

\citet{Gaudi:2009pmc} describes the \emph{separational} class of event
as typically occurring in archetypal binary microlenses with
resonant caustic crossings. If the binary's orbit is inclined, the
projected separation of the lenses changes, causing a stretching or
compression of the resonant caustic. If the projected separation is
close to a boundary between caustic topologies, $d \sim \dc$, or $d
\sim \dw$, the changes in the caustic structure can be very rapid. If
the microlensing event occurs while the changes are happening, and the source
crosses, or passes close to, the caustics, there is a very good chance
of detecting the orbital motion. As a whole though, the changes in
caustic structure during the caustic crossing time-scale will be
fairly small, e.g. the difference in caustic crossing time between the
static lens and the orbiting lens may be of order minutes to hours
(cf. the orbital period of several years). It is only the extremely
good accuracy with which caustic crossings can be measured and timed
that facilitates the high orbital motion detection probability. These
changes to the caustic shape will often be more significant than the
changes in orientation of the caustic due to rotation, and so we class
them as separational orbital motion effects.  

\citet{Gaudi:2009pmc} described the \emph{rotational} class of event
as occurring when a source encounters two disjoint caustics of a
typically close topology lens. In the time between the two caustic
encounters, which are separated by a time $\Delta t \sim \tein$, the
lens components have time to rotate and show detectable signatures of
orbital motion. We extend the class by considering the important
effect to be the long baseline over which binary lensing features can
be detected. If binary lens features are detectable across a
significant fraction of the lightcurve then a significant amount of rotation
can occur in the lens while the features are detectable. Up to now, our
discussion has focused mainly on caustic features, whether the source
crosses them or not, but, in stellar binary lenses especially, the
magnification pattern of the lens can differ significantly, if
subtly, from the single lens form over large parts of the pattern,
and well away from caustics. For example, in close binary lenses,
there is a region of excess magnification that can stretch the entire
distance between the facing cusps of the central and secondary
caustics. In stellar binary lenses, this can extend for distances
larger than an Einstein radius. In planetary lenses the magnification
excesses are weaker, but there tends to be a large region of
demagnification between the two planetary caustics. If lenses with
such features rotate rapidly, then the source may encounter them in
such a way that a static lens interpretation of the lightcurve
features is not possible, and lens rotation must be invoked.

%
%
\begin{figure}
\includegraphics[width=84mm]{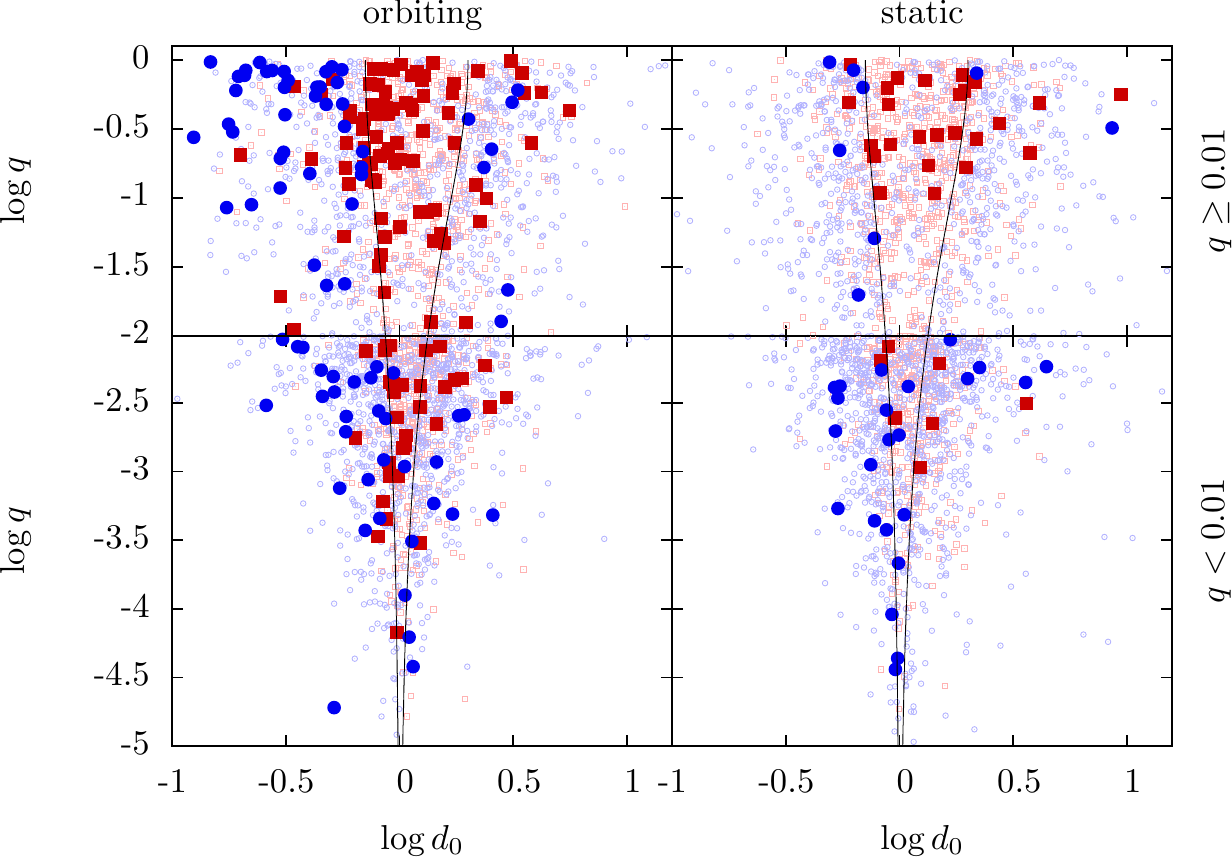}
\caption{Scatter plot of $q$ against $\dzero$ for microlensing events
  with detectable binary signatures. Caustic crossing events are
  plotted with red squares, and smooth events with blue
  circles. Events classified as orbital motion events are plotted with
  larger, darker filled points, and those classified as static with
  smaller, lighter, open points. The black lines show the positions of
  the caustic topology boundaries.}
\label{d0_q-scatter}
\end{figure}

We begin by looking for evidence of two classes of event in the locations of the orbital motion events in the $\dzero$-$q$ plane. Figure~\ref{d0_q-scatter} plots $q$ against $\dzero$ for all binary events; events which do not show orbital motion signatures are plotted with small, open points with light colours, whereas those that do are plotted with large, filled points with darker colours. Caustic crossing events are plotted with red squares, and smooth events with blue circles. Upper panels show stellar binary lenses and lower panels show planetary lenses, while the left panels show orbiting lenses and the right panel show static lenses. The black lines show the boundaries between the caustic topologies (equations~\ref{dc} and \ref{dw}). It is immediately clear that caustic crossing and smooth orbital motion events reside in different regions of the $\dzero$-$q$ plane, with virtually all events within the intermediate topology regime being caustic crossing. Almost all smooth orbital motion events are located in the close topology region. This broadly reflects the underlying pattern for all binary events, and is not in itself evidence of two classes of orbital motion events, but is instead a result of different caustic sizes in the different caustic topologies.

Another feature of the plot is the clustering of caustic crossing
orbital motion events near the boundary of the close and intermediate
topologies. It is close to the topology boundaries that the changes in
projected separation cause the largest changes in the caustics. It is
however difficult to attribute this clustering to faster caustic
motions due to separational changes, as orbital velocity is inversely
correlated with $\dzero$, and so there should be more orbital motion
events at smaller values of $\dzero$ in any case. In support of the
existence of a separational class, there is a hint of clustering
against the resonant-wide boundary. However, the caustic size peaks at
both topology boundaries, as the single resonant caustic stretches
before splitting apart into central and secondary caustics, possibly
meaning that simply the increased size of the caustics causes the
increased density of detections. 

%
%
\begin{figure}
\includegraphics[width=84mm]{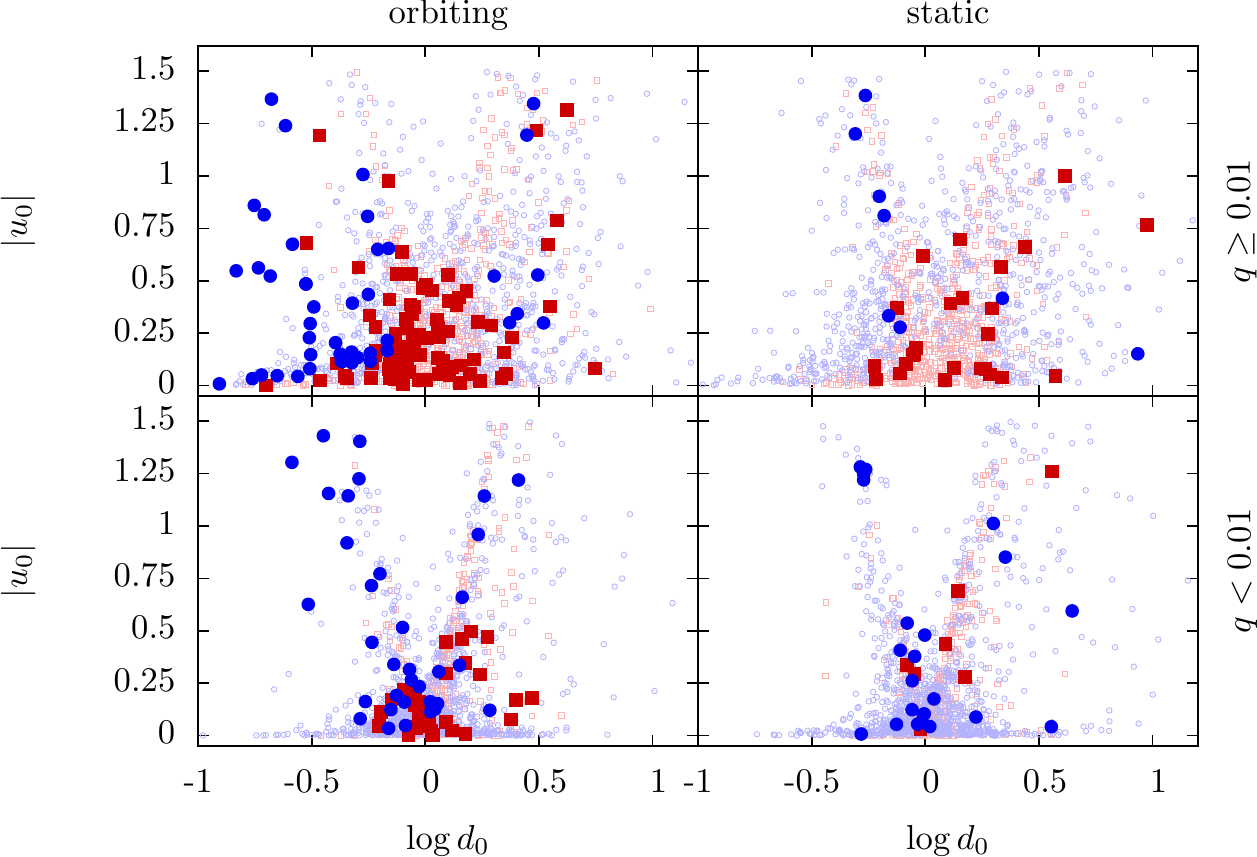}
\caption{As Figure~\ref{d0_q-scatter} but showing $|\uzero|$ plotted against $\dzero$.}
\label{d0_u0-scatter}
\end{figure}

Figure~\ref{d0_u0-scatter} plots the impact parameter against $\dzero$
and is very useful in separating different kinds of binary event,
especially for planetary lenses. The events follow a distinctive
pattern, with a large clump of events centred at $|\uzero| \sim 0$ and
$\log \dzero \sim 0$ which consists of high-magnification events that
encounter the central or resonant caustic. At very small $|\uzero|$
this clump extends over a significant range in $\dzero$, but narrows
as $|\uzero|$ increases, to its narrowest point at $|\uzero| \sim
0.3$, corresponding to the maximum size of the region affected by
resonant caustics (or at larger $|\uzero|$ for stellar binaries). As
$|\uzero|$ increases, the plot shows a distinctive `V' shape, with no
binary signatures being detected for events with $\dzero \sim 0$. This
`V' shape arises as, in events with larger $|\uzero|$, the source passes through regions of the
magnification pattern that can only contain secondary caustics, and
does not enter the regions containing central or resonant caustics,
i.e. the binary features in lenses with $\dzero \sim 1$ only occur in
regions of the magnification pattern that the sources with large
$|\uzero|$ do not probe. 

The events which occur on the branch with large $|\uzero|$ and large
$\dzero$ are caused by wide topology lenses, and therefore involve
only a single secondary caustic encounter. The rotation of these
lenses is typically very slow, and over the short duration of the
binary features (typically of order a day), the lens completes only a
very small fraction of its orbit. This points towards separational
changes being the dominant effect in the detection of orbital motion
features in events on this branch, even with the enhancement of
rotational velocity due to the solid body `lever arm'. 

The events that occur on the branch with large $|\uzero|$ and small
$\dzero$ are largely smooth events, with the occasional caustic
crossing event. The smooth events are likely caused by the source
crossing the large cusp extensions that occur in close binary lenses,
suggesting that they will belong to the rotational class of events.

\begin{figure*}
\rotatebox{270}{\includegraphics[width=210mm]{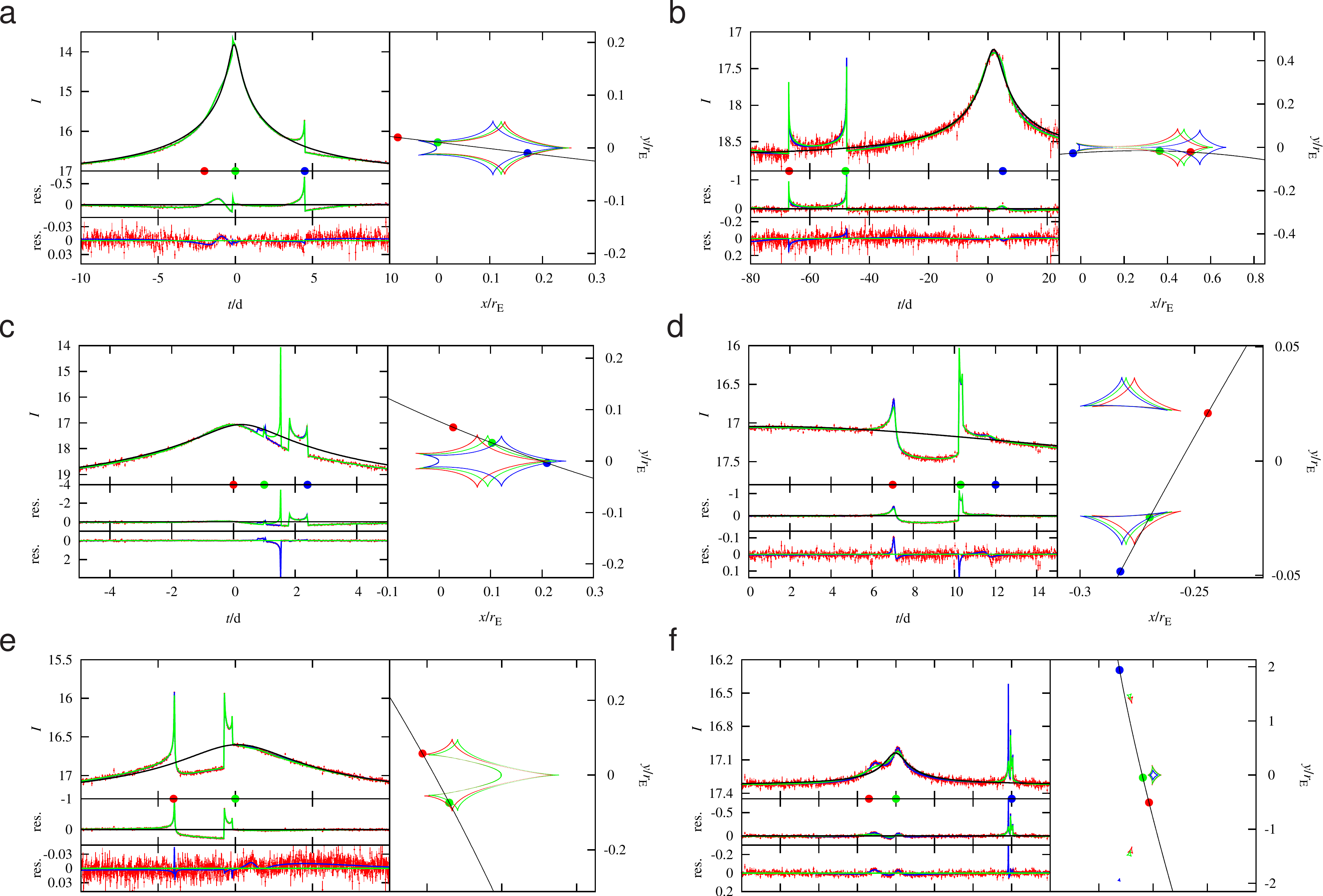}}
  \caption{Example lightcurves of simulated events affected by
  separational type orbital motion effects. In each subfigure, the
  left panels show the lightcurve, its residual with respect to the
  best fitting \paczynski model and its residual with respect to the
  best fitting static binary model, from top to bottom respectively. Simulated data is shown in red,
  the \paczynski model in black, the static binary model in green and
  the true model in blue. The right panel shows the caustics at
  various times, and the source trajectory in the frame of reference
  rotating with the projected binary axis. The source trajectory is
  plotted in black, and the caustics are colour coded according to the
  time. Coloured points on the lightcurve panel show the time at which
  the caustic was in the state shown, and the coloured points on the
  source trajectory show the position of the source at this time. The
  parameters of the microlensing events can be found in Tables~4 and 5
  in the online supplementary material.}
\label{separationalExamples}
\end{figure*}

\begin{figure*}
\rotatebox{270}{\includegraphics[width=210mm]{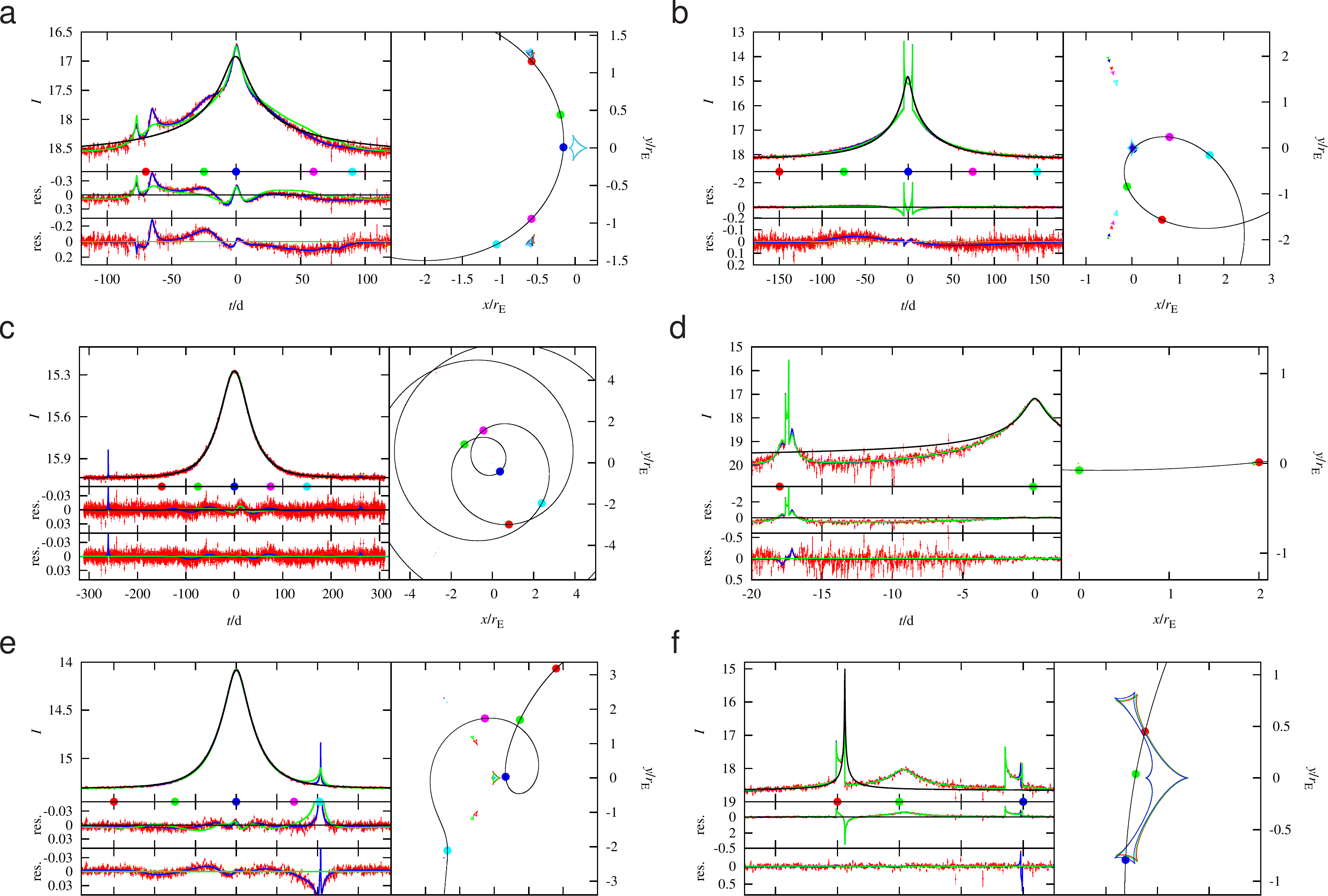}}
  \caption{As Figure~21, but showing example lightcurves of
  simulated events affected by rotational type orbital motion
  effects.}
\label{rotationalExamples}
\end{figure*}

\begin{table*}
\caption{Microlensing parameters of example lightcurves in the paper.}
\centering
\begin{tabular*}{168mm}{@{\extracolsep{\fill}}lcccccccc}
\hline
Figure & Orbit$^{\dag}$ & $\uzero$ & $\thetazero/\degr$ & $\dzero$ & $q$ & $\tein/$d & $\mzero$ & $\blendfs$ \\
\hline
3 & C & 0.48 & 307 & 8.64 & 0.22 & 14.9 & 17.9 & 1.04 \\
4 & S & -0.091 & 186 & 0.95 & 0.054 & 14.7 & 19.2 & 0.59 \\
5top & C & 1.43 & 315 & 5.23 & 0.030 & 7.5 & 18.8 & 0.41 \\
5middle & C & -0.16 & 155 & 0.61 & 0.14 & 12.6 & 19.3 & 0.082 \\
5bottom & C & 0.37 & 255 & 2.92 & 0.21 & 6.9 & 14.5 & 0.93 \\
21a & C & -0.011 & 255 & 1.06 & 0.0016 & 26.2 & 17.1 & 0.19 \\
21b & C & -0.024 & 285 & 1.31 & 0.0076 & 132.2 & 18.7 & 0.067 \\
21c & C & -0.071 & 81 & 1.04 & 0.0015 & 12.2 & 19.6 & 0.71 \\
21d & C & 0.22 & 265 & 0.87 & 0.00045 & 65.7 & 18.0 & 0.38 \\
21e & C & 0.16 & 169 & 0.94 & 0.0038 & 26.3 & 17.3 & 0.15 \\
21f & E & -0.20 & 16 & 0.55 & 0.49 & 14.8 & 17.3 & 0.073 \\
22a & C & 0.15 & 52 & 0.57 & 0.33 & 54.6 & 18.6 & 0.67 \\
22b & C & 0.033 & 69 & 0.45 & 0.56 & 88.3 & 18.2 & 0.72 \\
22c & C & -0.56 & 353 & 0.18 & 0.30 & 49.3 & 16.0 & 1.04 \\
22d & C & -0.076 & 245 & 2.38 & 0.0059 & 9.0 & 20.0 & 1.04 \\
22e & E & -0.33 & 163 & 0.34 & 0.29 & 82.4 & 15.3 & 0.96 \\
22f & E & 0.21 & 77 & 0.79 & 0.29 & 24.3 & 18.7 & 0.20 \\
\hline
\end{tabular*}
\\
\begin{tabular*}{168mm}{@{\extracolsep{\fill}}p{168mm}}
\small{$^{\dag}$C--circular orbit, S--static orbit, E--eccentric
  orbit}
\end{tabular*}
\label{microlensingParameters}
\end{table*}

\begin{table*}
\centering
\caption{Physical parameters of example lightcurves in the paper.}
\begin{tabular*}{168mm}{@{\extracolsep{\fill}}lccccccccc}
\hline
Figure & Orbit & $\bigmone/\msun$ & $\bigmtwo$ & $a/$AU & $T/$d & $e$ & $i/\degr$$^{\ddag}$ & $\vt/$km s$^{-1}$ & $\dl/$kpc \\
\hline
3 & C & 0.084 & 0.018~$\msun$ & 10.7 & 39799 & 0 & 214 & 134.8 & 5.75 \\
4 & S & 0.70 & 0.038~$\msun$ & 1.88 & 1090 & 0 & 300 & 215.7 & 7.40 \\
5top & C & 0.058 & 0.0018~$\msun$ & 4.46 & 14047 & 0 & 173 & 196.3 & 6.04 \\
5middle & C & 0.13 & 0.017~$\msun$ & 1.22 & 1298 & 0 & 311 & 183.8 & 5.95 \\
5bottom & C & 0.10 & 0.021~$\msun$ & 3.52 & 6852 & 0 & 112 & 282.8 & 6.43 \\
21a & C & 0.55 & 0.89~$\mjup$ & 5.82 & 6924 & 0 & 93 & 167.3 & 6.12 \\
21b & C & 0.75 & 6.0~$\mjup$ & 4.32 & 3767 & 0 & 115 & 39.8 & 6.01 \\
21c & C & 0.27 & 0.43~$\mjup$ & 0.51 & 256 & 0 & 243 & 63.2 & 7.91 \\
21d & C & 0.89 & 0.42~$\mjup$ & 3.83 & 2899 & 0 & 136 & 88.8 & 2.13 \\
21e & C & 1.17 & 4.7~$\mjup$ & 3.42 & 2130 & 0 & 56 & 173.5 & 7.19 \\
21f & E & 0.21 & 0.10~$\msun$ & 0.61 & 306 & 0.92 & 102,216 & 183.0 & 6.90 \\
22a & C & 0.56 & 0.18~$\msun$ & 1.88 & 1098 & 0 & 16 & 101.2 & 2.44 \\
22b & C & 0.38 & 0.21~$\msun$ & 1.69 & 1044 & 0 & 40 & 57.4 & 2.69 \\
22c & C & 0.68 & 0.20~$\msun$ & 0.65 & 205 & 0 & 30 & 115.8 & 5.97 \\
22d & C & 0.65 & 4.0~$\mjup$ & 2.70 & 2005 & 0 & 2 & 218.3 & 7.75 \\
22e & E & 0.59 & 0.17~$\msun$ & 1.35 & 656 & 0.77 & 303,213 & 68.2 & 5.56 \\
22f & E & 0.39 & 0.11~$\msun$ & 2.14 & 1609 & 0.18 & 2,143 & 187.0 & 5.64 \\
\hline
\end{tabular*}
\\
\begin{tabular*}{168mm}{@{\extracolsep{\fill}}p{168mm}}
\small{$^{\ddag}$For events with eccentric orbits, two values of
  inclination are quoted, representing inclinations about two
  orthogonal axes on the sky. The effect of this second inclination is
  absorbed into the source trajectory for circular orbits, and to
  first order can be reduced to the range $0\degr\le i \le 90\degr$.}
\end{tabular*}
\label{physicalParameters}
\end{table*}

Unfortunately it is difficult to attribute the cause of any one
grouping of orbital motion events in Figures~\ref{d0_q-scatter} and
\ref{d0_u0-scatter} to either the rotational or the separational
class, partly because both types of motion will affect each event to
some extent. Despite this, it is possible to classify many individual
events as either a separational or rotational
event. Figures~\ref{separationalExamples} and \ref{rotationalExamples}
show example lightcurves of both classes of orbital motion event,
rotational and separational, respectively. The plots show the
lightcurve in the upper left panels, with simulated data in red, the
true model in blue, the best fitting static binary model in green and
the best fitting single lens model in black. Also shown are the
residuals from the single lens model and the static binary model in
the middle and lower left panels respectively. Shown in the right
panel is a plot of the source trajectory, shown in black, and
snapshots of the caustics at various times during the event, shown in
different colours. The coloured points on the time axis of the
lightcurve show the time at which the caustic snapshots occurred, and
the coloured points on the source trajectory show the position of the
source at these times. The source trajectory and caustics are shown in
the frame of reference that rotates with the binary axis, with its
origin at the centre of mass. In this frame, rotation of the lens
causes the source trajectory to appear curved, and changes in lens
separation cause the caustics to change shape and move. Note that in
event f in Figure~\ref{separationalExamples}, and events e and f in
Figure~\ref{rotationalExamples}, the lens orbits are eccentric, so
that the source does not travel along the shown trajectory at a constant
rate. 

Figure~\ref{separationalExamples} shows examples of separational
events. In each example the source trajectory appears relatively
straight, indicating that the lens rotates little; however, in each
case the caustics move significantly. Events a, b, c and e all involve
resonant caustic crossings, and conform well to the picture described
by \citet{Gaudi:2009pmc}. Event d could be described as the encounter
of two disjoint caustics, similar to the original description of the
rotational class of events by \citet{Gaudi:2009pmc}, but other than
the close topology, the event is remarkably similar to event e; the
source trajectory is slightly curved, but it is clear that
separational effects are dominant. At first glance, event f would
clearly fit into the picture of disjoint caustic encounters, but the
source trajectory reveals that rotation plays only a minor role. In
this event, a static fit to just the features about $t=\tzero$ would suggest
a close encounter with a large secondary caustic at $t\approx
1.5\tein$, but instead changes in the binary's separation cause the
source to not just encounter, but cross a now much smaller secondary
caustic at $t\approx 2\tein$.

In contrast to Figure~\ref{separationalExamples}, the source
trajectories in Figure~\ref{rotationalExamples} show significant
curvature. Event a fits the description of rotational events by
\citet{Gaudi:2009pmc} exactly. The source first encounters a secondary
caustic, but the rotation of the lens causes the source to pass the
opposite side of the central caustic. Rotation also prevents the
source from crossing the magnification excess between central caustic
and the other secondary caustic. During the entire event, separational
changes cause only slight changes in the caustics. In event c the
rotation is more extreme, but the caustics smaller. The binary
features are therefore more subtle, being caused by small
magnification excesses between the caustics; the secondary caustics
being located at $\sim(-3,\pm 4)$, and the central caustic at $\sim
(0,0)$. The rotation of the lens causes the source to cross each
excess more than once, and there are several minor deviations visible
in the residual between the static and true model of the event. Event
d, while being caused by a wide lens, expected to rotate slowly, is
clearly caused by rotation. During the event, there are virtually no
separational changes, but the precision with which the secondary caustic
crossing and cusp approach features constrain the source trajectory
mean that the very slight rotation which brings the source closer to
the central caustic is detectable. Events b and e both show strong
signs of rotation in their source trajectories, but separational
changes are also important. While we assign them to the
rotational class of event, in reality they may better fit into a
third, hybrid class. Event f also shows signs of both rotational and
separational orbital motion effects, but we assign it to the
rotational class because without rotation the second caustic crossing
would be significantly shorter.

\section{Discussion}
\label{Discussion}

\subsection{Limitations of the study}

The questions that we wanted to answer in this work were: what
fraction of microlensing events observed by the next generation
surveys will be affected by orbital motion and what type of events are
the effects likely to be seen in? While we do not claim to have fully
answered these questions, we do feel that this work represents an
important step in that direction. The simulation of the photometry is
slightly optimistic, and does not include the effects of weather and
the systematic differences in the site conditions and observing systems, 
distributed across the Globe, that would make up the network of
telescopes needed for a continuous monitoring microlensing survey. The
observing setup we simulated is in some respects more like a space based microlensing
telescope than a ground based network. However, the photometric
accuracy that we simulated is not too optimistic, and the differences
between the static and orbiting simulations show that orbital motion
plays a significant role in a significant fraction of microlensing events.

As discussed in Section~\ref{Method}, our choice of
models will not fully answer the question of how many microlensing
events with orbital motion effects will be seen, however, they do
provide a good order of magnitude estimate. The binary detection
efficiencies we find assume that all stars have a companion, and so
must be adjusted accordingly to account for this. For example, current estimates suggest that only $\sim 33$~percent of stellar systems are binaries~\citep[e.g.][]{Lada:2006bsf}, so assuming that a next
generation microlensing survey detects $\sim 2000$ events per year we
can expect to see $\sim 30$ stellar binary microlensing events showing
orbital motion signatures per year. However, the true rate may be
higher as the mass ratio distribution that we use for stellar binaries
is not realistic; the real distribution is likely to be peaked in
the range $0.1\le q \le 1$~\citep[e.g.][]{Duquennoy:1991bsp}. A
similar calculation for planetary lenses, assuming the fraction of
stars hosting planets is $\sim 0.5$, yields a detection rate of $\sim
1.5$ caustic crossing orbital motion events per year. Again, this
estimate is affected significantly by our assumptions. Our mass ratio
distribution is optimistic, with current microlensing results
suggesting an inverse relation between planet frequency and mass ratio
in the regions microlensing is sensitive
to~\citep{Sumi:2010nps,Gould:2010pps}. This implies our estimate is
optimistic, but we have also assumed there is only one planet per
system. Many multiplanet systems have been discovered to
date~\citep[e.g.][]{Gaudi:2008jsa,Fischer:2008fps}, and they are
thought to be common. The microlensing planet detection efficiency in
multiplanet systems is increased, as the planets are spread
over a range of semimajor axes. This will somewhat compensate for the
overestimate due to the incorrect mass ratio distribution.

The major limitation of this work is that finite source effects are
not considered. The finite size of the source acts to smooth out the
extreme magnification peaks as a source crosses a caustic, limiting the
precision with which magnifications can be measured, and caustic
crossings timed, and thus plays an important role in orbital motion detection. However, in most cases, the caustic entry times can still be
timed accurately if the caustic crossing is monitored with high enough
cadence. In some cases the effect may increase the
detectability of orbital motion as the source will probe more of the
magnification pattern, especially when a source travels approximately
parallel to and very close to the inside of a fold caustic, producing
additional peaks between the caustic crossings. We cannot quantitatively
estimate the effects that finite source size has on the orbital motion
detection efficiency, but we do not believe it will significantly
affect our order of magnitude estimates. Unfortunately including
finite source sizes in the modelling of a microlensing event increases
the required computation time by several orders of magnitude, so the
effect could not easily be included in the simulations without
significantly reducing the sample size. 

\subsection{Comparison with observations}

While our simulations are more representative of future microlensing
surveys, it is possible for us to compare the results of our
simulations with the results of the current microlensing
observations. Current microlensing planet searches using the
survey/follow-up strategy routinely achieve a cadence similar or better
than that expected for future high-cadence surveys for a small number
of microlensing events per year~\citep[e.g.][]{Dong:2009bjp}. We can
therefore compare the detection efficiency of orbital motion in the
events where planets are detected. At the time of writing, there were 10
published detections of planets by microlensing~\citep{Bond:2004pml, Udalski:2005jmp, Beaulieu:2006fem, Gould:2006cnp, Gaudi:2008jsa, Bennett:2008lmp, Dong:2009bjp, Sumi:2010nps, Janczak:2010ssp}, and of these, 7 had high cadence coverage of a significant proportion of the
lightcurve. In two of these events the orbital motion of the planet was
detected~\citep{Gaudi:2008jsa,Dong:2009mmm},\footnote{While the orbital
  motion of the Jupiter analogue was not detected in the
  OGLE-2006-BLG-109 system, the planet itself would still have been
  detected in the absence of the Saturn analogue, so it contributes
  to the denominator of the detection efficiency, but not to the
  numerator.} leading us to estimate an orbital motion detection
efficiency of $\sim \ase{0.29}{0.10}{0.13}$~percent. This efficiency is larger
than we find in our simulations. However, the orbital motion effects
in the OGLE-2005-BLG-71 event are very subtle, and improve the fit by
$\dcstat\ll 200$~\citep{Udalski:2005jmp,Dong:2009mmm},\footnote{The
  overall reduction in $\chi^2$ between the two analyses was much less
  than $200$ when the size of the data sets and differing degrees of
  freedom are accounted for. The full analysis by \citet{Dong:2009mmm}
  included higher order effects not included in the original
  \citet{Udalski:2005jmp} analysis, some of which had a much larger
  effect than orbital motion.} meaning that it would not be classed as a
detection in our simulations; this reduces the comparable detection
efficiency estimate to $\ase{0.14}{0.07}{0.11}$, with which our estimates of the detection efficiency for planetary caustic crossing events are
consistent. It should be noted that this figure could be biased as events showing orbital motion signatures will take significantly longer to analyze. Unfortunately a similar estimate for binary star
lenses is not so simple as they are usually not followed-up to the
same degree that planetary events are, either in terms of observations
or modelling.

We have identified two different classes of orbital motion event so it
is natural to try to classify the orbital motion events that have
already been seen. The orbital motion detected in
OGLE-2006-BLG-109~\citep{Gaudi:2008jsa,Bennett:2010jsa} was detected
due to deformation of a resonant caustic, and the event can easily be
assigned to the class with separational
changes. OGLE-2005-BLG-71~\citep{Udalski:2005jmp,Dong:2009mmm} is
harder to classify, as the orbital motion effects observed were very
subtle. The event suffers from the well known close-wide
degeneracy~\citep{Griest:1998cwd,Dominik:1999cwd}, and rather
strangely, for the close ($d<1$) solution, separational changes are
more prominent than rotational, and vice versa for the wide ($d>1$)
solution, where we might normally expect the opposite. We therefore do not assign the event to either class. Of
the stellar binary lenses, MACHO-97-BLG-41~\citep{Albrow:2000rbl} was
mainly influenced by rotation, and was detected by two disjoint
caustic crossings, so is classed as a rotational
event. EROS-2000-BLG-5~\citep{An:2002eb5} undoubtedly belongs to the
separational class; the caustic structure was resonant with $d$ close
to $\dw$, and changes in separation were measured with high
significance, while rotational changes were consistent with zero. The
final events, OGLE-2003-BLG-267 and
OGLE-2003-BLG-291~\citep{Jaroszynski:2005bme} are not very well
constrained, so we do not attempt to classify them.

We finally suggest that the event
OGLE-2002-BLG-069~\citep{Kubas:2005blo} is a strong candidate for
showing rotational type orbital motion effects. The event was modelled
successfully by \citet{Kubas:2005blo} without including orbital
motion, with a close binary solution favoured physically and by the
modelling. The event had a time-scale $\tein\approx 105\tdayunit$ and
binary parameters $d=0.46$ and $q=0.58$. The lightcurve was very
similar to event b shown in Figure~\ref{rotationalExamples}, having a
long, well covered central caustic crossing, with measurements of both caustic
entry and exit. The physical lens parameters obtained from the
modelling suggest lens masses of $\bigmone=0.51\msun$ and
$\bigmtwo=0.30\msun$, and a projected separation of $\sim 1.7$~AU,
with a corresponding minimum period of $T\gtrsim 900\tdayunit$. The
baseline is relatively bright, at $\mzero\sim 16.2$, and so subtle magnification deviations could probably be constrained by the data, if they have been covered.

\subsection{Future prospects}

Interestingly, our results show that the orbital motion
detection efficiency depends only weakly on the mass ratio. In the case of planetary events, caustic crossing orbital motion detections occur preferentially in high to moderate magnification events ($A\gtrsim 5$), while smooth orbital motion detections occur in all but high-magnification events. 
Our results therefore suggest that the strategy of targeting
high-magnification events~\citep{Griest:1998cwd,Han:2001mol} should
allow caustic crossing orbital motion events to be detected
efficiently. However, the strong dependence of orbital motion
detection efficiency on the event time-scale suggests that long
time-scale events should also be routinely followed-up. While
follow-up of these events requires a significant investment of
resources from the follow-up teams, like high-magnification events
they are relatively rare. For a given cadence, these events allow a
better signal to noise detection of planetary deviations, and also
allow more time for the prediction of future features. Long time-scale
events are also more likely to show parallax features, allowing
constraints to be placed on the lens mass.

High cadence, continuous monitoring microlensing surveys will begin
operating in the next few years. Already, the MOA-II
survey~\citep{Hearnshaw:2005moa, Sako:2008moa} has been surveying a
fraction of its total survey area with a cadence of $\sim 10$ minutes
for some time, and the OGLE-IV survey~\citep{Udalski:2009og4} has
begun operations this year, and should provide significant increases
in cadence over OGLE-III. KMTnet, a uniform network of telescopes with near continuous coverage, and operating at a cadence of $\sim 10$ minutes should begin operating around 2013; this promises an almost order of magnitude increase in
the detection rate of microlensing events, and a similar if not bigger
increase in the detection rate of planets by microlensing. The uniform
nature of the survey network will also make statistical analysis of
the planets detected easier, greatly enhancing the work already done in
this direction~\citep{Sumi:2010nps,Gould:2010pps}. The work we have
presented shows that a significant fraction of the events will show
signs of orbital motion, which will significantly complicate the
interpretation of future planet detections. However, these
complications can be used to provide valuable additional constraints
on the lens.

Often overlooked are binary star microlensing events. The next
generation surveys will detect many more binary star events
than planetary events. A large number of these lenses will be located in the
Galactic bulge and be composed of low mass stars, providing an
opportunity to study the properties of the bulge binary star population. Our results show that a significant fraction of these events will show orbital motion signatures, and it is likely that in a significant
number of these events it will be possible to measure the masses of
the system. It should therefore be possible to measure the statistics of a population that is difficult to reach by current spectroscopic and astrometric
methods due to their low brightness and long periods. 

\section{Summary}
\label{Summary}

We have simulated the lightcurves of $\sim 100,000$ microlensing events
caused by stars orbited by a companion star or planet. By fitting
simulated data with single lens and static binary models we have
determined the fraction of these events where the binarity of the lens
is detected, and we also estimate the fraction of these events where
orbital motion is detected. For an observational set up that resembles
a near future microlensing survey conducted by a global network of
telescopes without intensive follow-up observations, we found that orbital motion was detected in $\sim
5$--$10$~percent of simulated binary star microlensing
events depending on the characteristics of the event. Similarly, the rate of detection of orbital motion in simulated microlensing events where a planet is detected was $\sim 1$--$5$~percent. 

We investigated the effects of various event parameters on the
fraction of events showing orbital motion. Orbital motion detection
efficiency as a fraction of binary detections was found to depend only weakly on the mass ratio of the binary, but strongly on the event time-scale. We found that a significant number of microlensing events showing orbital motion can be classified into one of two classes: those where the dominant cause of orbital motion effects is either the separational motion of the binary due to either inclination or eccentricity, or the rotational motion of the binary.

%
%
%
%

\section*{Acknowledgments} 
MP acknowledges the support of an STFC studentship and would like to
thank Scott Gaudi for several interesting and helpful discussions, Martin Smith for providing the blending data and invaluable suggestions regarding the false positives correction, and the anonymous referee for many suggestions that improved the paper.

%
%
%
%

\bibliographystyle{mn2e}
\bibliography{mn-jour,orbitalMotion.bib}

\end{document}